\documentclass[prl,aps,twocolumn,superscriptaddress]{revtex4-2}
\usepackage{graphicx,color,setspace}
\usepackage[utf8]{inputenc}
\usepackage{mathtools}
\usepackage[normalem]{ulem}
\usepackage{tabularx}
\usepackage{enumerate}%
\usepackage{float}

\usepackage[caption=false]{subfig} %
\usepackage{ragged2e} %
\DeclareCaptionJustification{justified}{\justifying}

\begin{document}
\title{Two-step nucleation in a binary mixture of Patchy Particles}

\author{Camilla Beneduce}
\thanks{Both authors contributed equally}
\affiliation{Dipartimento di Fisica, Sapienza Universit\`{a} di Roma, P.le Aldo Moro 5, 00185 Rome, Italy}
\author{Diogo E. P. Pinto}
\thanks{Both authors contributed equally}
\affiliation{Dipartimento di Fisica, Sapienza Universit\`{a} di Roma, P.le Aldo Moro 5, 00185 Rome, Italy}
\author{Petr \v{S}ulc}
\affiliation{School of Molecular Sciences and Center for Molecular Design and Biomimetics, The Biodesign Institute, Arizona State University, 1001 South McAllister Avenue, Tempe, Arizona 85281, USA}
\affiliation{Life and Medical Sciences Institute (LIMES), University of Bonn, Bonn, Germany}
\author{Francesco Sciortino}
\affiliation{Dipartimento di Fisica, Sapienza Universit\`{a} di Roma, P.le Aldo Moro 5, 00185 Rome, Italy}
\author{John Russo}
\affiliation{Dipartimento di Fisica, Sapienza Universit\`{a} di Roma, P.le Aldo Moro 5, 00185 Rome, Italy}

\begin{abstract}
Nucleation in systems with a metastable liquid-gas critical point is the prototypical example of a two-step nucleation process, in which the appearance of the critical nucleus is preceded by the formation of a liquid-like density fluctuation. So far, the majority of studies on colloidal and protein crystallization have focused on one-component systems, and we are lacking a clear description of two-step nucleation processes in multicomponent systems, where critical fluctuations involve coupled density and concentrations inhomogeneities. Here, we examine the nucleation process of a binary mixture of patchy particles designed to nucleate into a diamond lattice. By combining Gibbs-ensemble simulations and direct nucleation simulations over a wide range of thermodynamic conditions,
we are able to pin down the role of the liquid-gas metastable phase diagram on the nucleation process.
In particular, we show that the strongest enhancement of crystallization
occurs at an azeotropic point with the same stoichiometric composition of the crystal.
\end{abstract}

\maketitle

\section{Introduction}

The ability to guide dilute solutions of properly designed colloidal particles to
spontaneously turn into specific ordered structures is crucial for a large-scale application of  the self-assembly paradigm~\cite{whitelam2015statistical,kumar2017nanoparticle}. 
One relevant example of this strategy is provided by the DNA-origami nanotechnology, nowadays a mature methodology which allows the 
design and production of nanometric particles with controllable shape and interactions~\cite{seeman2017dna}. It has been demonstrated that these particles, once properly designed, are able to self-assemble into ordered crystalline lattices even starting from quite dilute conditions~\cite{liu2016diamond,zhang20183d}, suggesting the exploitation of complex nucleation pathways that are still not fully understood.

One such pathway is the so-called \emph{two-step nucleation pathway}, in which 
a metastable gas-liquid transition can spontaneously generate local regions with high particle concentration, thus greatly enhancing the nucleation rate~\cite{kashchiev2005kinetics,erdemir2009nucleation,vekilov2010two,toth2011amorphous,sear2012,haxton2015crystallization,russo2016nonclassical,sosso2016crystal,lutsko2019crystals,james2019phase,desgranges2019can,kashchiev2020classical}.
In the last decade, this process has been observed in simple systems that are either one-component systems, or ionic compounds: examples include
hard particle fluids~\cite{lee2019entropic}, colloids~\cite{ten1997enhancement,tan2014,haxton2015crystallization}, salt solutions~\cite{jiang2019nucleation}, calcium carbonate~\cite{gebauer2008,pouget2009}.
Importantly, two-step processes play an important role in the crystallization of globular proteins~\cite{ten1997enhancement,fusco2013crystallization,james2015self,mcmanus2016physics,fusco2016soft}.
For colloidal particles, globular proteins and non globular proteins, the short-range nature of the inter-particle interaction generates a metastable gas-liquid coexistence which provides a mechanism for locally enhancing the density and triggering crystal formation. Spinodal decomposition and/or liquid-phase nucleation are equally active mechanisms.  In addition, ten Wolde and Frenkel~\cite{ten1997enhancement} have called attention on the possibility to exploit critical fluctuations as a mechanism for locally increasing density. In all these cases, a two-step nucleation process takes place, the first step being the
local increase of density, and the second the formation of a crystal nucleus in one or several of such enhanced density regions.

The crystallization of specific crystal structures is not devoid of problems~\cite{jee2016nanoparticle,dijkstra2021predictive}. Kinetic pathways can drive the assembly towards amorphous aggregates, or crystals different from the desired one. To address this last case,
several strategies have been suggested, examples include: optimization of preparation protocols~\cite{whitelam2020learning,bupathy2022temperature,whitelam2021neuroevolutionary}, the accurate design of the interparticle potential~\cite{rechtsman2005optimized,marcotte2011optimized,marcotte2013designeddiamond,zhang2013probing,miskin2016turning,lindquist2016communication,chen2018inverse,kumar2019inverse,dijkstra2021predictive,kumar2019inverse}, or geometrical approaches
where
the inter-particle interactions are made to match the geometric features of the target structure~\cite{ducrot2017colloidal,nelson2002toward,manoharan2003dense,zhang2005self,romano2014influence,halverson2013dna,romano2012patterning,tracey2019programming}.
Rather than increasing the complexity of the inter-particle interactions, another successful strategy is to increase the number of building blocks.
When the building blocks are modeled as patchy colloids, i.e. particles with specific directional interactions, the process of designing the bonding rules to assemble one specific crystal (and selectively avoid the formation of any other competing structure)
 can be turned into a set of boolean satisfiability (SAT) equations, which can be solved using modern SAT solver algorithms. This strategy, named \emph{SAT-assembly}, was shown to lead to the successful assembly of several complex target structures, including the coveted colloidal diamond crystal~\cite{romano2020designing,russo2022sat,rovigatti2022simple}.

Compared to the one-component case, crystal formation and two-step pathways in multi-component systems have received much less attention. In these systems, composition fluctuations couple with density fluctuations and it is not a priori clear if the stochiometric properties of the 
spontaneously generated liquid phase are consistent with the corresponding properties of the selected crystal. Even when the components are fully miscible in the solid phase (such as Pd and Ag), it was shown that the competition between demixing and crystallization has a big impact on the growth process, causing large variations in the radial composition of the nuclei~\cite{desgranges2014unraveling}.
Multi-component systems also display thermodynamic behaviour that is not found in one-component systems, such as lines of binary critical points, azeotropic points, re-entrant condensation, etc. Experimental studies have already shown that phase separation in multi-component systems does not preserve the original ratio between the different species and that the phase behaviour strongly depends on the components ratio ~\cite{wang2010phase,wang2011phase,heidenreich2020designer}, but it is yet unclear whether some of the peculiar multi-component mixture features can affect crystal nucleation.

In this manuscript we use extensive computer simulations of a binary mixture of patchy particles that self-assembles in a diamond crystal with fixed stoichiometric ratio, with the goal of relating the phase behaviour of the mixture with its crystallization pathway.
The manuscript is organized as follows.
In the Methods section we introduce the N2c8 binary mixture and briefly summarize its phase behaviour. Numerically, we obtain the phase diagram with Gibbs ensemble simulations, while the crystallization behaviour is observed via direct Monte Carlo simulations. We make use of biased moves to accelerate diffusional times and observe crystallization for state points where the free-energy barrier can be overcome by thermal fluctuations during our simulation times. The Results section will superimpose the metastable phase diagram with histograms of the nucleation rate, confirming the two-step nucleation pathway and allowing us to study the relation between demixing and crystallization. In the final section we put our results in the wider context of Empty Liquids, and argue that while the two-step nucleation pathway is an effective strategy to overcome diffusional barriers, azeotropic points are the key to lower compositional barriers.

\section{Methods}

We choose patchy particles as our model system: they describe systems with a hard-core isotropic repulsion and attractive directional interactions. Experimentally, the anisotropic interactions can be obtained either via the shape~\cite{van2013entropically} or via chemical patterning of the surface~\cite{zhang2004self,pawar2010fabrication,bianchi2011patchy,romano2011colloidal,suzuki2009controlling}, for example by attaching single strands of DNA to well-defined positions on their surface~\cite{suzuki2009controlling,kim2011dna,wang2012colloids,feng2013dna,Rothemund2006,tian2020ordered}.
We represent them computationally via the 
Kern-Frenkel~\cite{bol1982monte,kern2003fluid} potential, in which hard-core spherical particles of diameter $\sigma$ interact with a square well potential $V_{SW}$ of depth $\epsilon$ and width  $\delta$, modulated by a term $f$ depending on the relative orientation of the particles: in particular, two patchy particles can establish a \emph{bond} if their centers are at distance between $\sigma$ and $\sigma+\delta$, and if the line connecting their centres intersects the volume of both patches involved in the bond.
In the following we measure energy in units of the square-well depth ($\epsilon$) and distances in units of the patchy particle diameter ($\sigma$).
Mathematically, the  interaction between particle $i$ and $j$ is described by

\begin{equation}
	\label{eqn:KF}
	V(\mathbf r_{ij},\hat{\mathbf r}_{\alpha,i},\hat{\mathbf r}_{\beta,j})= V_{SW}(r_{ij})f(\mathbf r	_{ij},\hat{\mathbf r}_{\alpha,i},\hat{\mathbf r}_{\beta,j})
\end{equation}

\noindent where  $\hat{\mathbf r}_{\alpha,i}$ ($\hat{\mathbf r}_{\beta,j}$) indicates the position of patch $\alpha$ ($\beta$) of particle $i$ ($j$), 
\noindent and

\begin{equation}
	\label{eqn:f_KF}
	f(\mathbf r_{ij},\hat{\mathbf r}_{\alpha,i},\hat{\mathbf r}_{\beta,j})=
	\begin{cases}
		\Upsilon_{(\alpha,i)\,(\beta,j)} &\text{if}\quad 
		\begin{array}{l}
			\hat{\mathbf r}_{ij} \cdot \hat{\mathbf r}_{\alpha,i} > \cos{(\theta_{max})}\\ 
			\hat{\mathbf r}_{ji} \cdot \hat{\mathbf r}_{\beta,j} > \cos{(\theta_{max})}
		\end{array} \\
		0 &\text{otherwise}
	\end{cases}
\end{equation}

$\Upsilon_{(\alpha,i)\,(\beta,j)}$ is the (color) interaction matrix element and is equal to $1$ if patch $\alpha$ on particle $i$ interacts with patch $\beta$ on particle $j$, and $0$ otherwise. $\mathbf{\Upsilon}$ is a matrix which encodes the interactions between both identical and distinct species. Finding the interaction matrix that will guide the system to self-assemble into a desired structure can be cast into a \emph{inverse self-assembly} problem, whose solution can be found with the \emph{SAT-assembly} algorithm~\cite{romano2020designing,russo2022sat}.

We run two types of simulations: Gibbs-ensemble simulations to determine the liquid-gas coexistence line, and NVT Monte Carlo simulations in an extensive range of densities ($\rho$) and relative concentrations ($x$, i.e. the fraction of species 2).

We fix the parameters of the Kern-Frenkel potential to $\cos{(\theta_{max})}=0.98$ and $\delta=0.2$, for which nucleation is readily accessible within simulation times.
Moreover, direct simulations take advantage of AVB biased moves~\cite{rovigatti2018simulate} to accelerate the dynamics of bond-formation and bond-breaking. Both methods have been described in more detail in Ref.~\cite{beneduce2022include}. We confirmed with independent simulations that nucleation events are observed also in unbiased simulations, albeit requiring longer simulation times.

\begin{figure}[!t]
    \centering
    \includegraphics[width=0.45\textwidth]{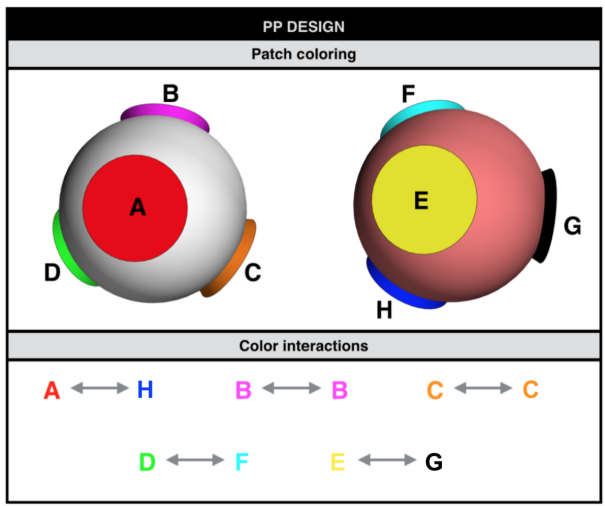}
    \caption{3D representation of the azeotropic N2c8 solution SAT-designed to self-assembly a cubic diamond crystal lattice. It consists of two patchy particles species (N2), the grey and the red one differing in the patch types. The patch types/different colors are eight (c8) referred to by the letters A through H. Patches that can bind with each other are reported in the bottom part of the table.}
    \label{fig_pp}
\end{figure}

Our goal is to study the nucleation process in multi-component mixtures such as the binary mixture described by the N2c8 interaction matrix (so called because it has 2 species and 8 colored patches) which was originally introduced as a SAT-assembly solution to the problem of assembling the cubic diamond crystal, while at the same time avoiding the hexagonal diamond crystal ~\cite{rovigatti2022simple}.
The ability to suppress a competing polytype, thus avoiding the formation of stacking faults, makes this mixture a candidate for the experimental realization of photonic crystals.
The N2c8 solution is graphically depicted in Fig.~\ref{fig_pp}: two patchy particles species differing in the patch types (colors); patches with matchable colors that can bind with each others are reported in the bottom part of the table. This mixture crystallizes in the diamond-cubic phase, and notably cannot form the hexagonal diamond.

\begin{figure}[!t]
    \centering
    \includegraphics[width=0.45\textwidth]{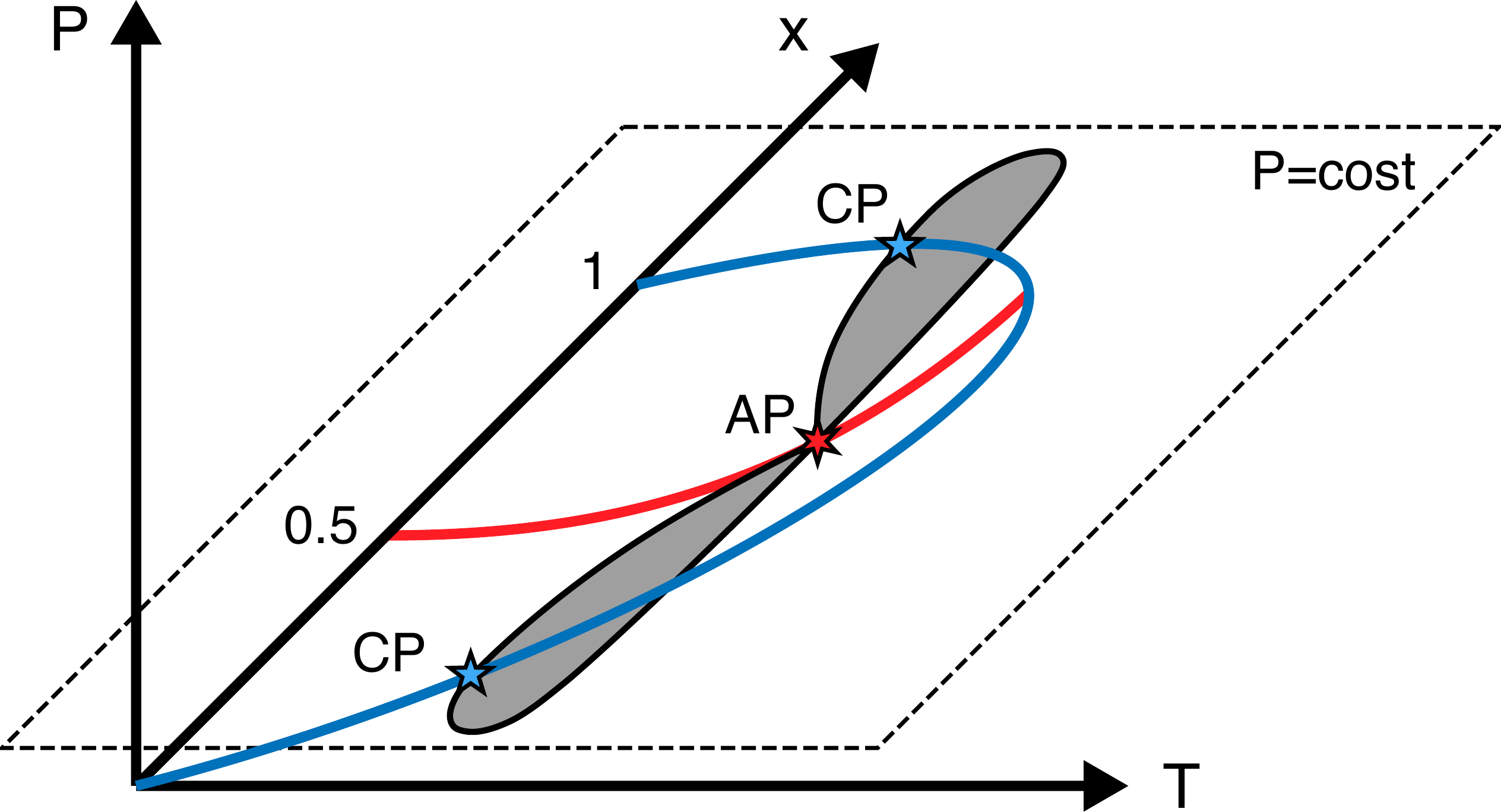}
    \caption{Schematic $PTx$ phase diagram of the N2c8 solution. The blue line is the binary critical point line which goes to $(P,T)\rightarrow 0$ for $x\rightarrow (0,1)$. The red line, in the plane $x=0.5$, represents the line of azeotropic points. The shaded area represents the coexistence region on a isobaric surface.}
    \label{fig_schematic}
\end{figure}

We highlight that the results presented in the following section for the N2c8 solution are general; we focus on this binary mixture since it is a meaningfull example and since we already know its phase behaviour. Indeed the phase diagram of the mixture was studied within Wertheim thermodynamic perturbation theory in Ref.~\cite{beneduce2022include}, which results we summarize in Fig.~\ref{fig_schematic}. The N2c8 mixture has a reentrant binary critical point line, which goes to zero temperature and pressure as the pure components are approached. This behaviour is immediately apparent from the color design of Fig.~\ref{fig_pp}, where it can be seen than each species can form only two intra-species bonds, so when the system is made of only one component it can only form linear chains, and thus does not have a critical point at any finite temperature~\cite{rubinstein2003polymer}.
The non-ideal nature of the mixture manifests itself in a line of azeotropic points at $x=1/2$. Along such line the system always demixes at the same composition, i.e. the liquid phase will  have an equimolar composition of both species. As we will see, this is particularly relevant for the nucleation process, as the crystal also has an equimolar composition.
In the results section we will look at the interplay between phase separation and nucleation.

To identify crystal particles we use local
bond-order analysis~\cite{tanaka2019revealing}. A
$(2l+1)$ dimensional complex vector ($\mathbf{q}_l$) is defined for each
particle $i$ as $q_{lm}(i)=\frac{1}{N_b(i)}\sum_{j=1}^{N_b(i)}
Y_{lm}(\mathbf{\hat{r}_{ij}})$, where we set $l=12$, and $m$ is an integer that
runs from $m=-l$ to $m=l$. The functions $Y_{lm}$ are the spherical harmonics
and $\mathbf{\hat{r}_{ij}}$ is the normalised vector from particle $i$ to particle $j$.  The sum goes over the first
$N_b(i)=16$ neighbours of particle $i$. We then introduce a
spatial coarse-graining step
$Q_{lm}(i)=\frac{1}{N_b(i)}\sum_{k=0}^{N_b(i)}q_{lm}(k)$~\cite{lechner2008accurate}.
The scalar product between $Q_{12,m}$ of two particles
is defined as $\mathbf{Q}_{12}(i)\cdot\mathbf{Q}_{12}(j)=\sum_m Q_{12,m}(i)Q_{12,m}(j)$.
If the scalar product
$(\mathbf{Q}_{12}(i)/|\mathbf{Q}_{12}(i)|)\cdot(\mathbf{Q}_{12}(j)/|\mathbf{Q}_{12}(j)|)$
between two neighbours exceeds $0.75$ then the two particles are deemed
\emph{connected}. We then identify particle $i$ as crystalline if it is
connected with at least $12$ neighbours.

\section{Results}

\begin{figure}[!t]
    \centering
    \includegraphics[width=0.45\textwidth]{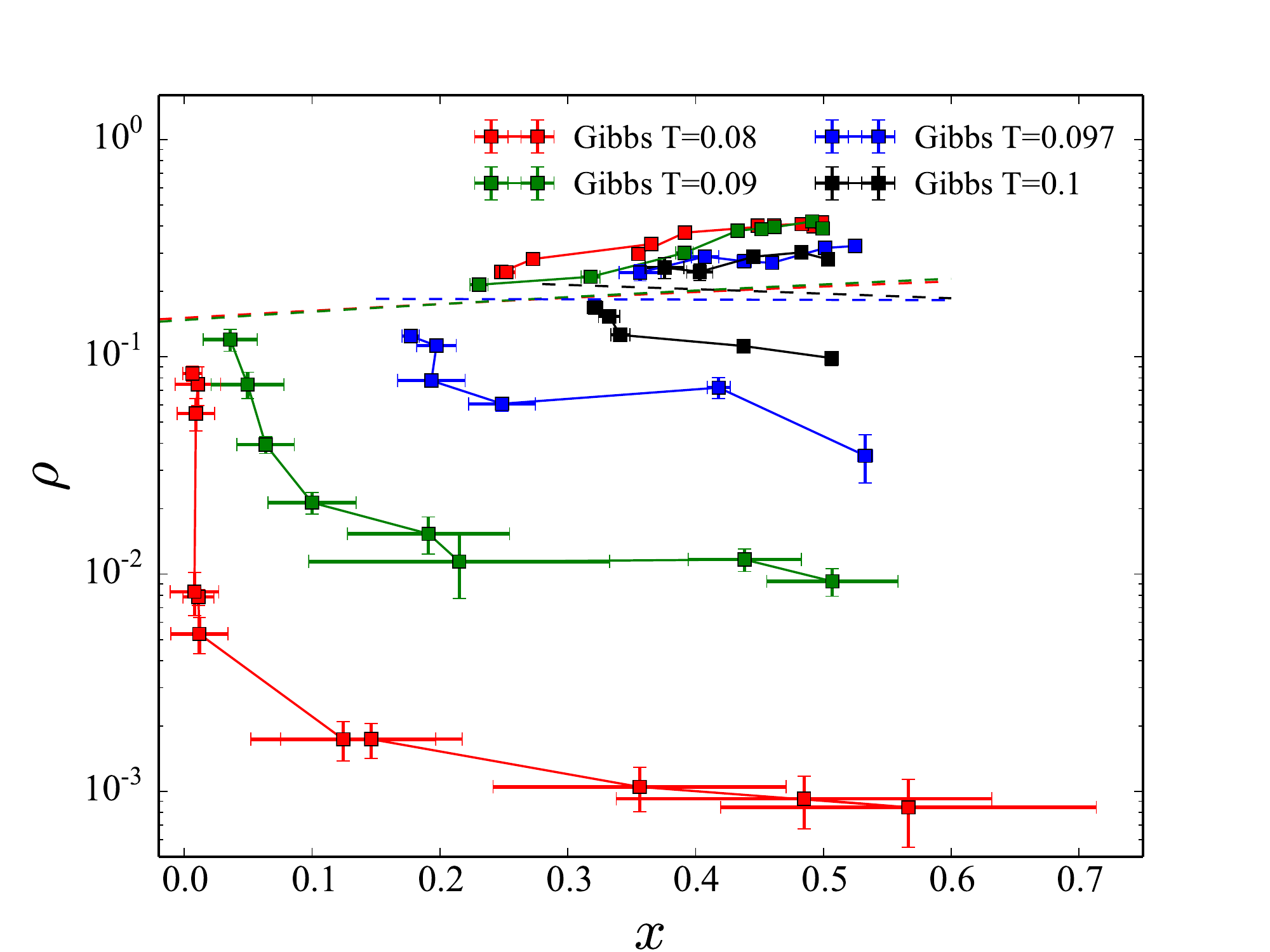}
    \caption{SAT-designed N2c8 binary mixture density-concentration phase diagrams for different temperatures $T$ expressed in unit of $k_{B}/\epsilon$;  the coexistence region shrinks by increasing temperature. Each binodal curve is obtained by Monte Carlo simulations with AVB moves in the Gibbs ensemble~\cite{rovigatti2018simulate}. In particular for temperatures $T=0.08$ and $T=0.09$, $500$ particles were simulated while for temperatures $T=0.097$ and $T=0.1$, $1000$ particles were considered. Dashed lines are rectilinear diameter lines indicating where critical points are located.}
    \label{fig_Gibbs_all_T}
\end{figure}

We run Gibbs ensemble simulations to determine the coexistence region at different temperatures. These are plotted in Fig.~\ref{fig_Gibbs_all_T} for temperatures $T=0.08$ (red), $T=0.09$ (green), $T=0.097$ (blue), and $T=0.1$ (black). The binodal line at each temperature is splitted in a liquid branch, at high densities, and a gas branch, at low densities. The two branches meet at the binary critical point, which location can not be estimated with Gibbs ensemble simulations~\cite{frenkel2001understanding}. The phase diagram is plotted only for $x<0.5$, but it is symmetric with respect to the azeotropic line $x=0.5$.  The critical point is located at the intersection of the binodal line and the rectilinear diameter line (i.e. the line connecting the midpoint of the different tie-lines), which is represented for each temperature as a dashed line in Fig.~\ref{fig_Gibbs_all_T}. For temperatures $T=0.08$ and $T=0.09$ we run single trajectories, while for temperatures $T=0.097$ and $T=0.1$ simulations are averaged over at least 8 independent simulations; the statistical noise is due to the difficulty in equilibrating the mixtures at low temperatures.

\begin{figure}[!t]
    \centering
    \includegraphics[width=0.5\textwidth]{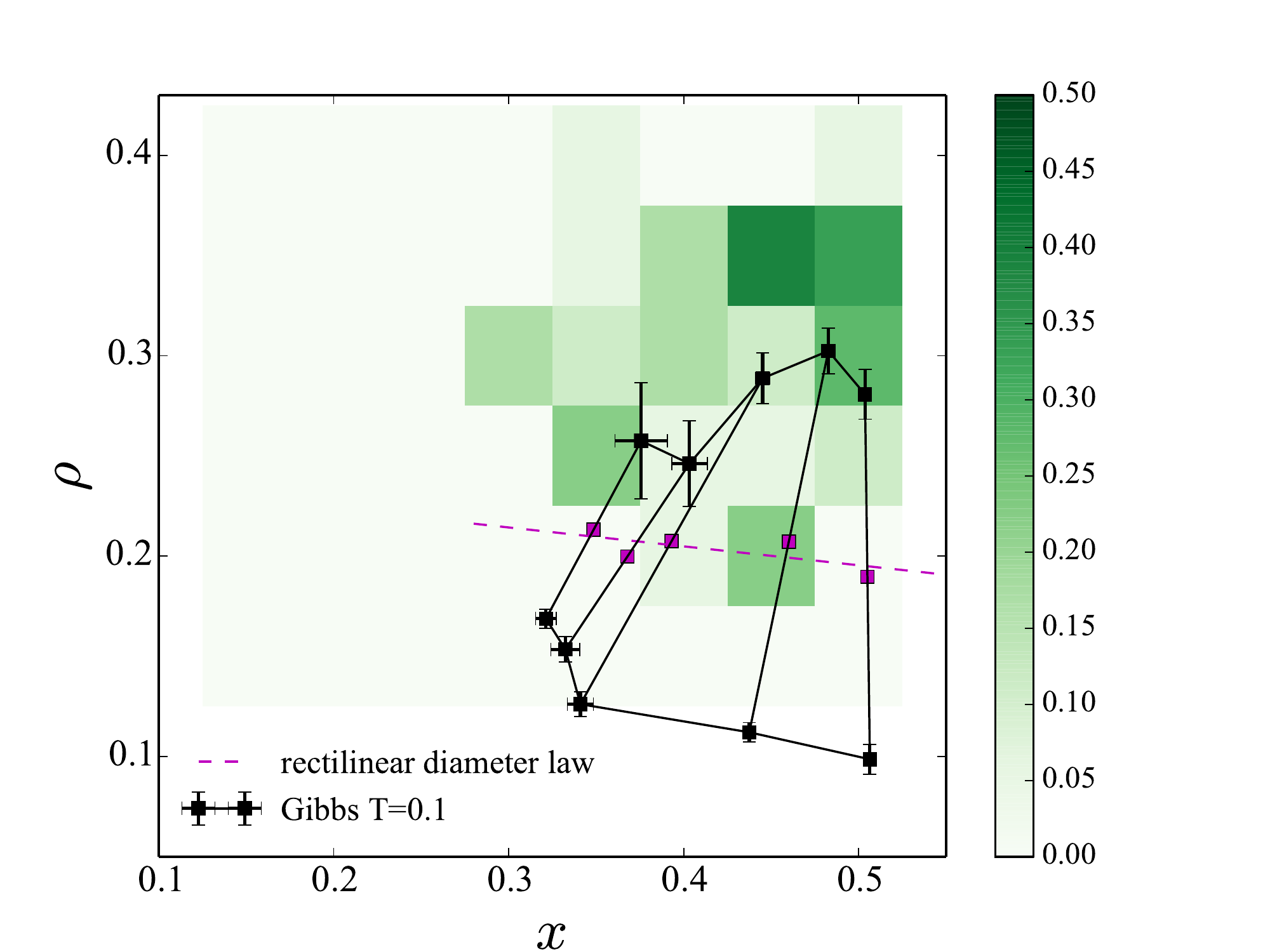}
    \caption{Nucleation plot for temperature $T=0.1$ (in unit of $k_{B}/\epsilon$). The binodal black curve obtained via Monte Carlo simulations in the Gibbs ensemble is superimposed to a green colormap reporting, for each considered state point, the ratio between the number of trajectories that nucleate over the total number of trajectories analysed. We considered as successfully assembled the runs exhibiting a fraction of particles in the cubic diamond phase greater or equal to $0.2$. Examined state points belong to the regular grid with $\rho \in [0.15, 0.4]$ and $x \in [0.15,0.5]$ where $\Delta\rho=0.05$ and $\Delta x=0.05$. For each state point (centred in each rectangular of the colormap) several Monte Carlo simulations were run with $500$ particles and AVB dynamics in the NVT ensemble. In particular we made $10$ runs for each state point whose $x\leq0.25$ and we increased them to $18$ for state points with $x\geq0.3$. Purple squares represent averages of each pair of coexisting points and, according to the rectilinear diameter law, the best straight line passing through them (the dashed line) provides an indication of where the critical point is.}
    \label{fig_nucl_01}
\end{figure}

In Fig.~\ref{fig_nucl_01} we examine the nucleation behaviour by running simulation runs over a grid of state points at $T=0.1$: $\rho\in[0.15,0.4]$ with $\Delta\rho=0.05$, and $x\in[0.15,0.5]$ with $\Delta x=0.05$. We run up to 10 (or 18 for state points where nucleation is more probable) independent simulation runs for more than $3\cdot 10^8$ steps each, and measure the fraction of trajectories that have nucleated, which is a quantity proportional to the nucleation rate. In the figure we define a nucleation event as a run in which at least $20\%$ of the particles are found in a crystalline state, according to the $Q12$ criteria described in the Methods section. 
In Fig.~\ref{fig_nucl_01} this fraction is proportional to the color intensity of a tile centered around the state point, and the grid is superimposed on the $T=0.1$ phase diagram. The figure shows that nucleation occurs in correspondence of the liquid-gas coexistence region, with the probability of nucleation being approximately the highest along the liquid binodal. We note in fact that no nucleation events are observed at low concentrations, from $x=0.15$ to $x=0.25$, i.e. outside the phase coexistence region. Visual inspection of the nucleation trajectories confirm that the nucleation event is preceded by the formation of a dense liquid phase, mostly via spinodal decomposition (i.e. liquid drops do not nucleate but instead a coarsening liquid networks appears at the start of the simulations).

\begin{figure*}[!t]
\centering
 \subfloat[\label{fig:T0097_02}]{
  \includegraphics[width=0.47\textwidth]{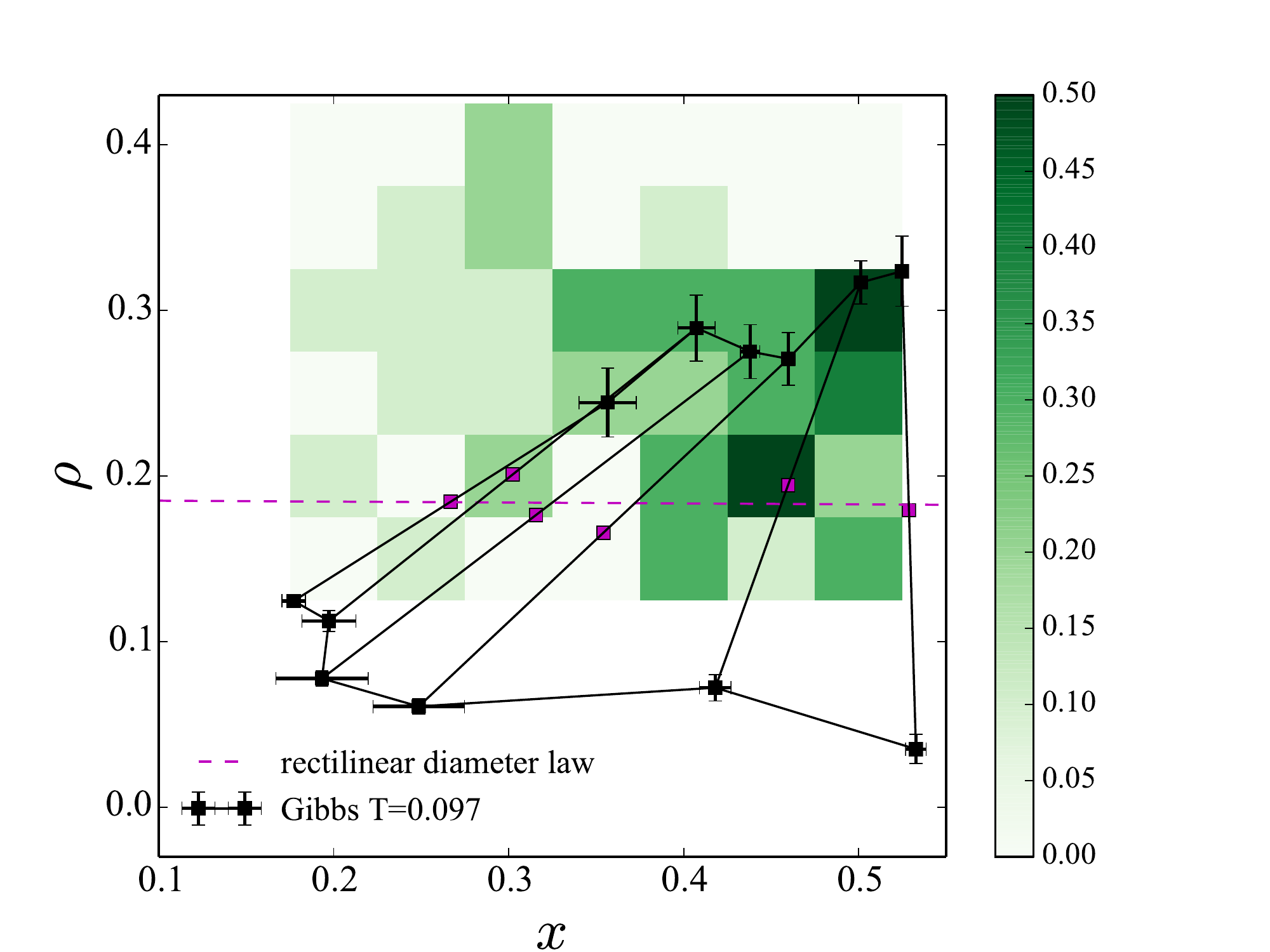}
  }
  \subfloat[\label{fig:T0097_05}]{
  \includegraphics[width=0.47\textwidth]{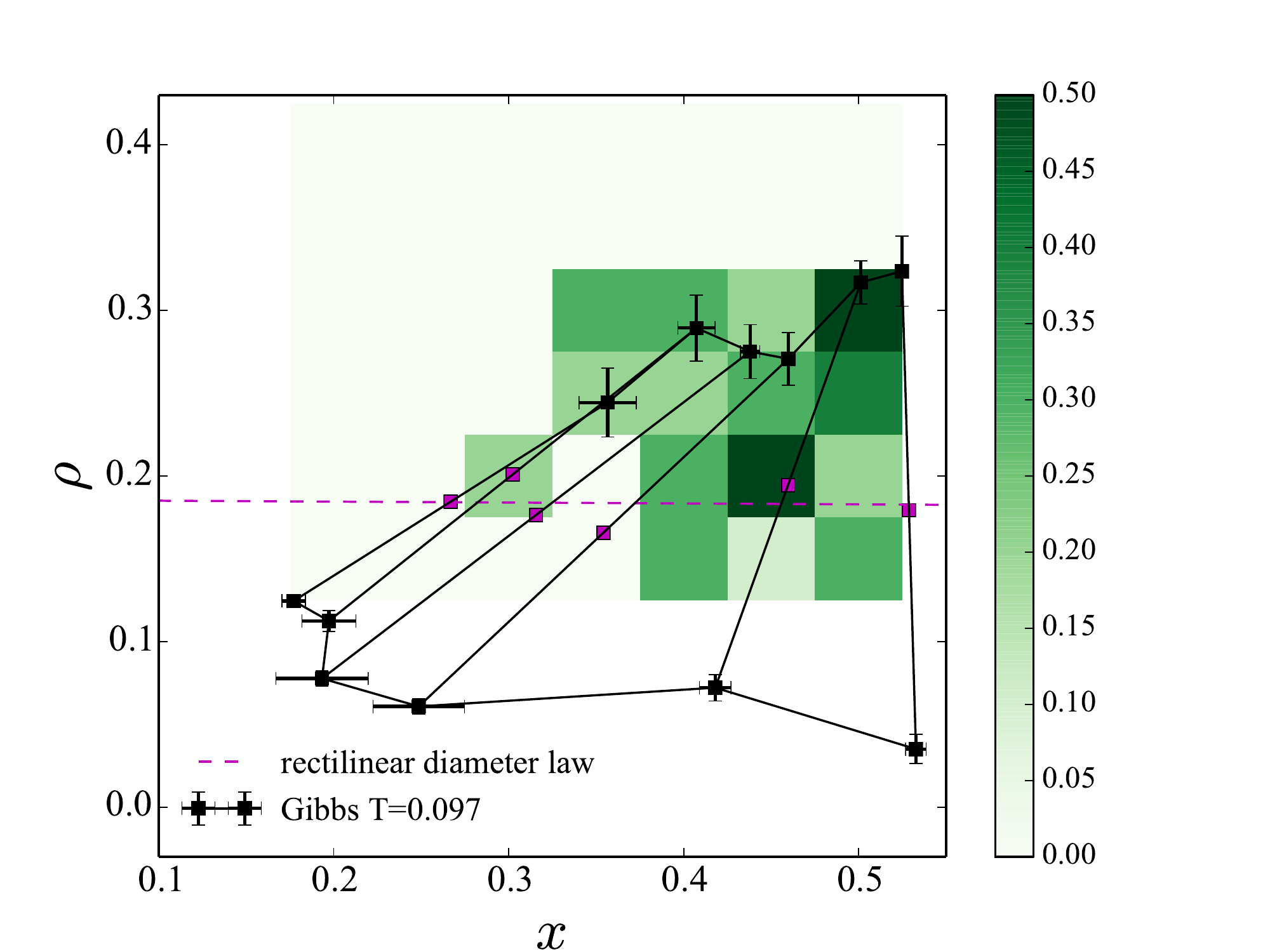}
  }
  \vskip\baselineskip
  \subfloat[\label{fig:T0097_07}]{
  \includegraphics[width=0.47\textwidth]{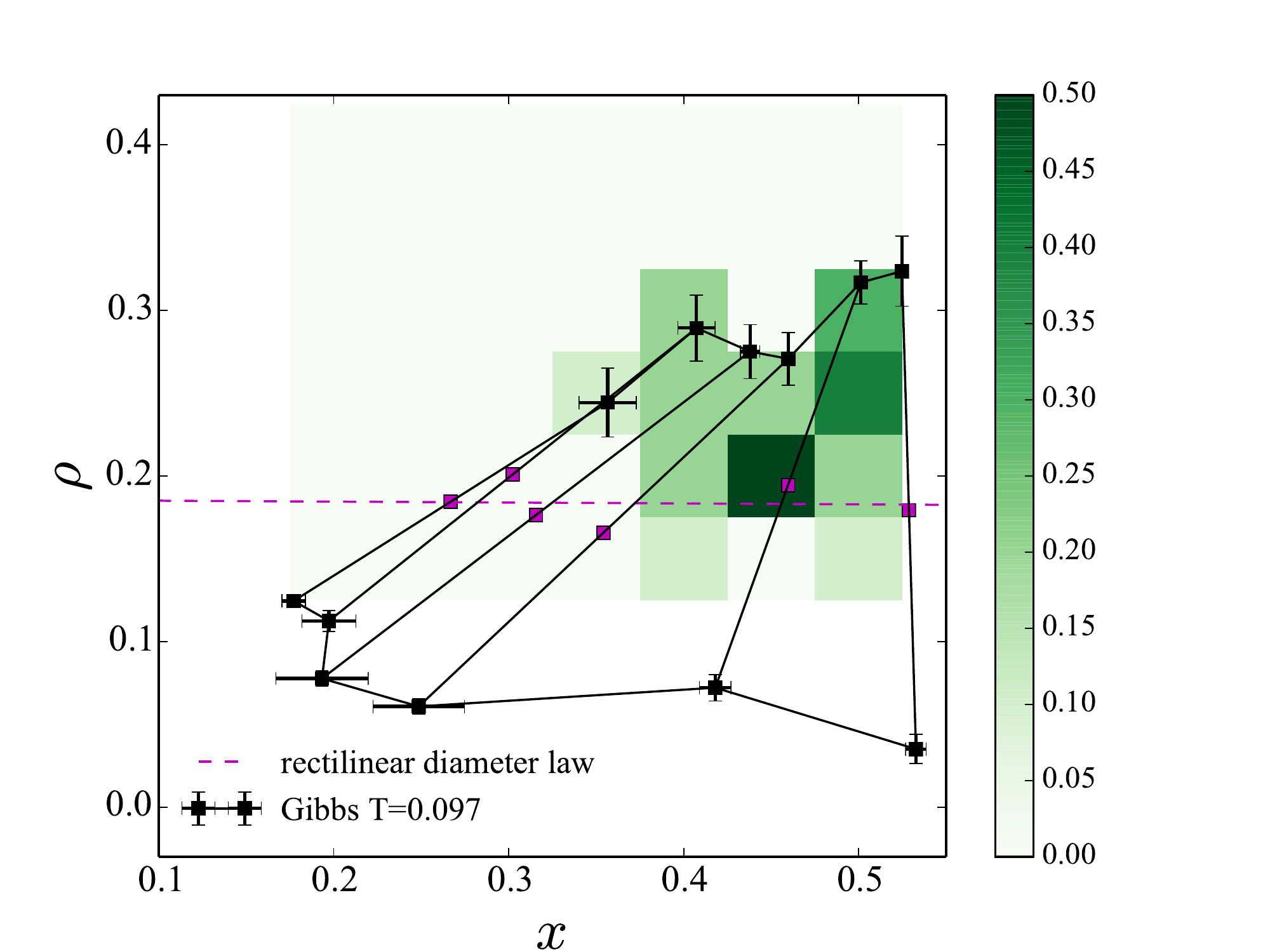}
  }
  \subfloat[\label{fig:T0097_09}]{
  \includegraphics[width=0.47\textwidth]{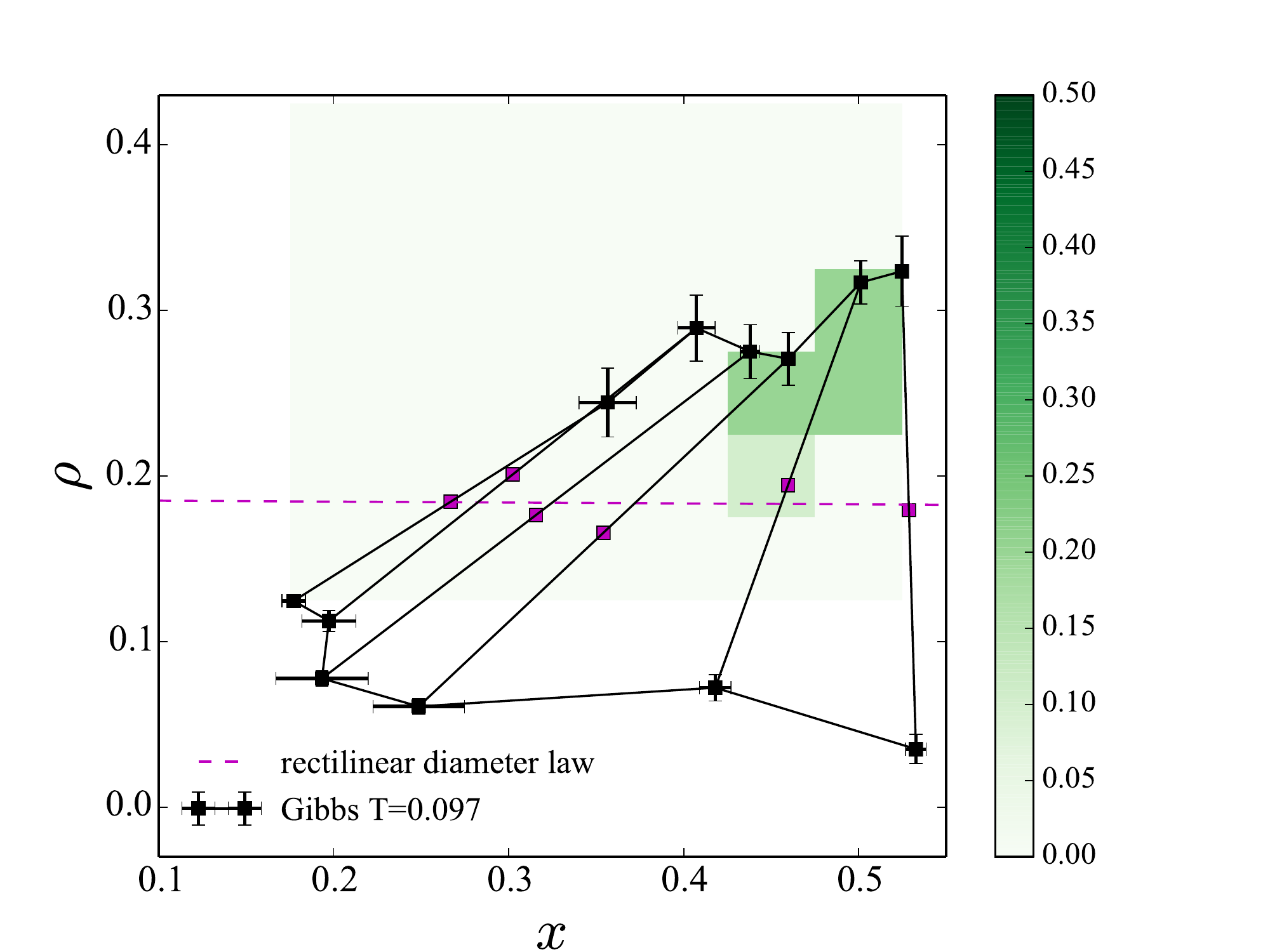}
  }
  \caption{Nucleation plots for temperature $T=0.097$ (in unit of $k_{B}/\epsilon$). The binodal black curve obtained via Monte Carlo simulations in the Gibbs ensemble is superimposed to a green colormap reporting, for each considered state point, the ratio between the number of trajectories that nucleate over the total number of trajectories analysed. We considered as successfully assembled the runs exhibiting a fraction of particles in the cubic diamond phase ($t$) greater or equal to $0.2$ in a), $0.5$ in b), $0.7$ in c) and $0.9$ in d).  Examined state points belong to the regular grid with $\rho \in [0.15, 0.4]$ and $x \in [0.2,0.5]$ where $\Delta\rho=0.05$ and $\Delta x=0.05$. For each state point (centred in each rectangular of the colormap) were run $10$ Monte Carlo simulations with $1000$ particles and AVB dynamics in the NVT ensemble. Purple squares represent averages of each pair of coexisting points and, according to the rectilinear diameter law, the best straight line passing through them (the dashed line) provides an indication of where the critical point is. Nucleation occurs for almost each considered state point, but extended cubic diamond lattice ($t\geq0.7$) can appear only close to the liquid branch and to the azeotropic condition.}
  \label{fig_nucl_0097}
\end{figure*}

To confirm the results of Fig.~\ref{fig_nucl_01}, we run nucleation trajectories in a density-concentration grid at a lower temperature, $T=0.097$, which makes the coexistence region wider. The results are presented in Fig.~\ref{fig_nucl_0097}(a). Compared to the case at $T=0.1$ now nucleation events occur over a much wider region of the phase diagram, following the widening of the coexistence region.

These results confirm the observations that were made for one-component systems, i.e. that density fluctuations promote nucleation~\cite{ten1997enhancement,xu2012homogeneous,haxton2015crystallization,james2019phase}. But compared to one-component systems, in a multi-component system there are both density and concentration fluctuations. Since the crystal has a fixed stoichiometric composition, the formation of a liquid phase at concentrations different from equimolarity is expected to slow down and limit the growth of the crystal. We show this effect in the four panels of Fig.~\ref{fig_nucl_0097}, where the colored grid plots the fraction of independent trajectories whose crystalline particles include at least the $20\%$ (a), $50\%$ (b), $70\%$ (c), and $90\%$ (d) of the whole system. Fig.~\ref{fig_nucl_0097} shows that, while small crystals can form along coexistence regions, the formation of larger crystals is progressively shifted towards the $x=0.5$ line, i.e. towards the azeotropic point. At the azeotropic point, in fact, the liquid phase forms with the same stoichiometric composition of the crystal.

\begin{figure}[!t]
    \centering
    \includegraphics[width=0.45\textwidth]{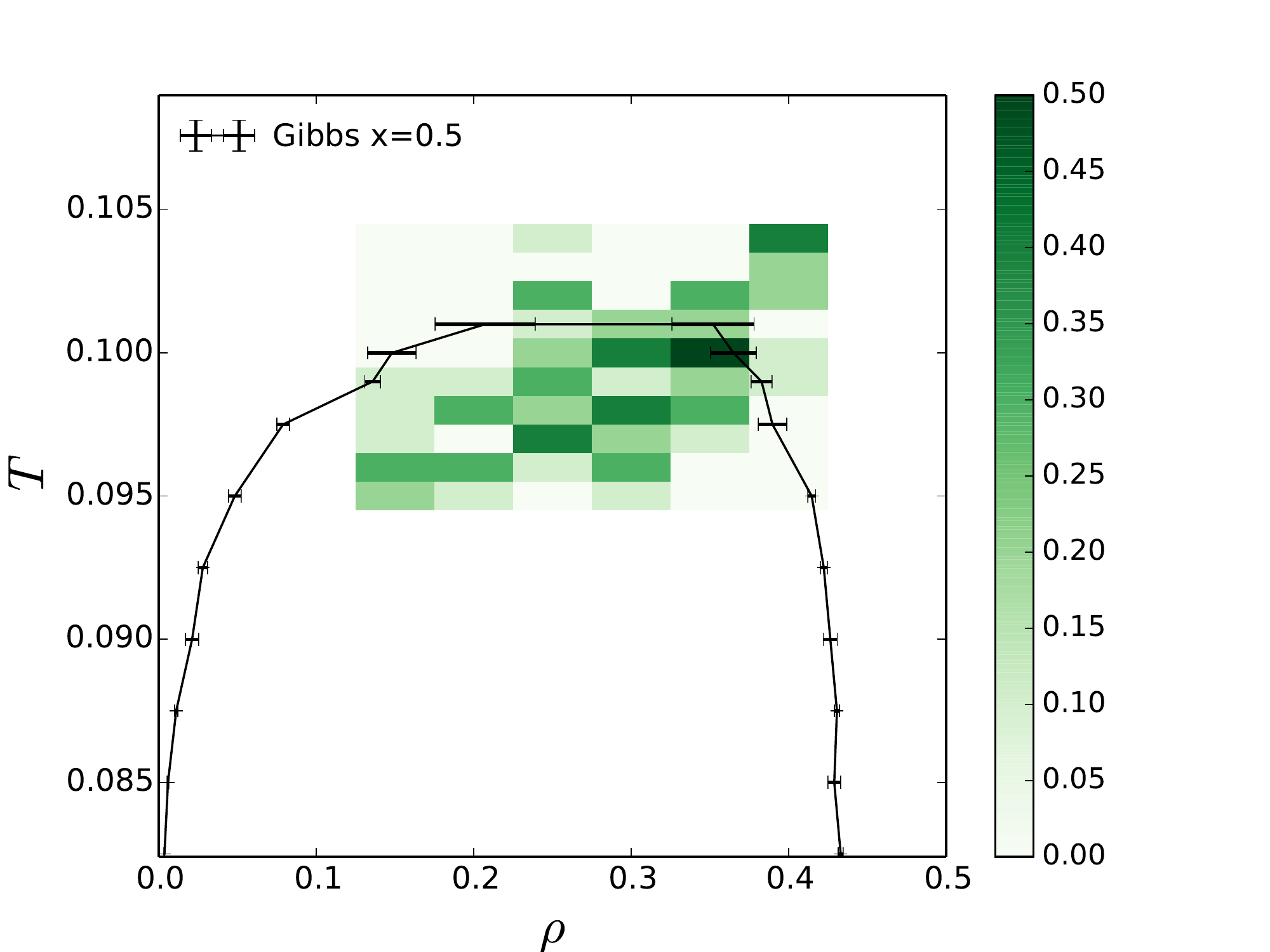}
    \caption{Equimolar nucleation plot showing that nucleation occurs in two steps where there is a previous locally increasing density phenomenon. The binodal black curve obtained via Monte Carlo simulations in the Gibbs ensemble is superimposed to a green colormap reporting, for each considered state point, the ratio between the number of trajectories that nucleate over the total number of trajectories analysed. We considered as successfully assembled the runs exhibiting a fraction of particles in the cubic diamond phase greater or equal to $0.2$. Examined state points belong to the regular grid with $\rho \in [0.15, 0.4]$ and $T \in [0.095,0.105]$ where $\Delta\rho=0.05$ and $\Delta T=0.001$. For each state point (centred in each rectangular of the colormap) $10$ Monte Carlo simulations were run with $500$ particles and AVB dynamics in the NVT ensemble. }
    \label{fig_nucl_equi}
\end{figure}

As demonstrated originally in Ref.~\cite{ten1997enhancement}, critical fluctuations are able to trigger crystal nucleation. We find here that in a binary mixture, it is the azeotropic point which more strongly promotes crystallization.
The azeotropic point and the critical point coincide where the two corresponding thermodynamic lines meet (see Fig.~\ref{fig_schematic}), which corresponds to the maximum critical temperature at $x=0.5$. In Fig.~\ref{fig_nucl_equi} we study the nucleation behaviour in the density-temperature plane at $x=0.5$.
The nucleation of the diamond cubic occurs again in correspondence of the phase separation, confirming that nucleation is aided by the formation of dense liquid regions during the phase-separation process. We observe that nucleation not only occurs in correspondence of the critical point, but, interestingly, we observe nucleation events at supercritical conditions. The system thus shows two-step nucleation aided both by spinodal decomposition at sub-critical temperatures, and by critical fluctuations at super-critical conditions, just as predicted by Ref.~\cite{ten1997enhancement}. While the prediction was made for isotropic one-component systems (globular proteins), at the azeotropic conditions of Fig.~\ref{fig_nucl_equi} the system is indeed thermodynamically equivalent to a one component system~\cite{beneduce2022include}.

 \begin{figure*}
    \centering
    \subfloat[\label{fig:energy_0097}]{
    \includegraphics[width=0.33\textwidth]{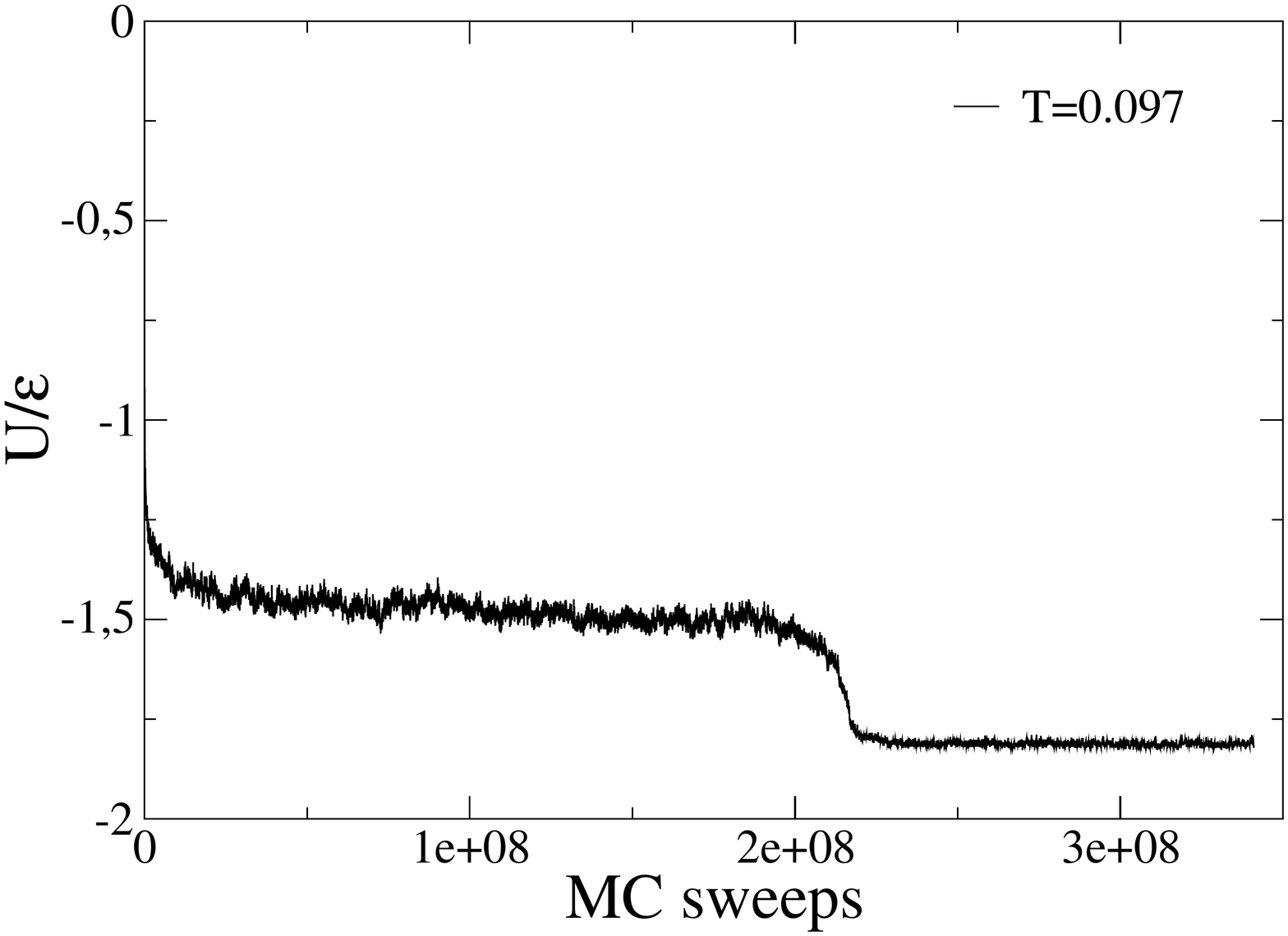}
    }
    \subfloat[\label{fig:box_0097_mid}]{
    \includegraphics[width=0.33\textwidth]{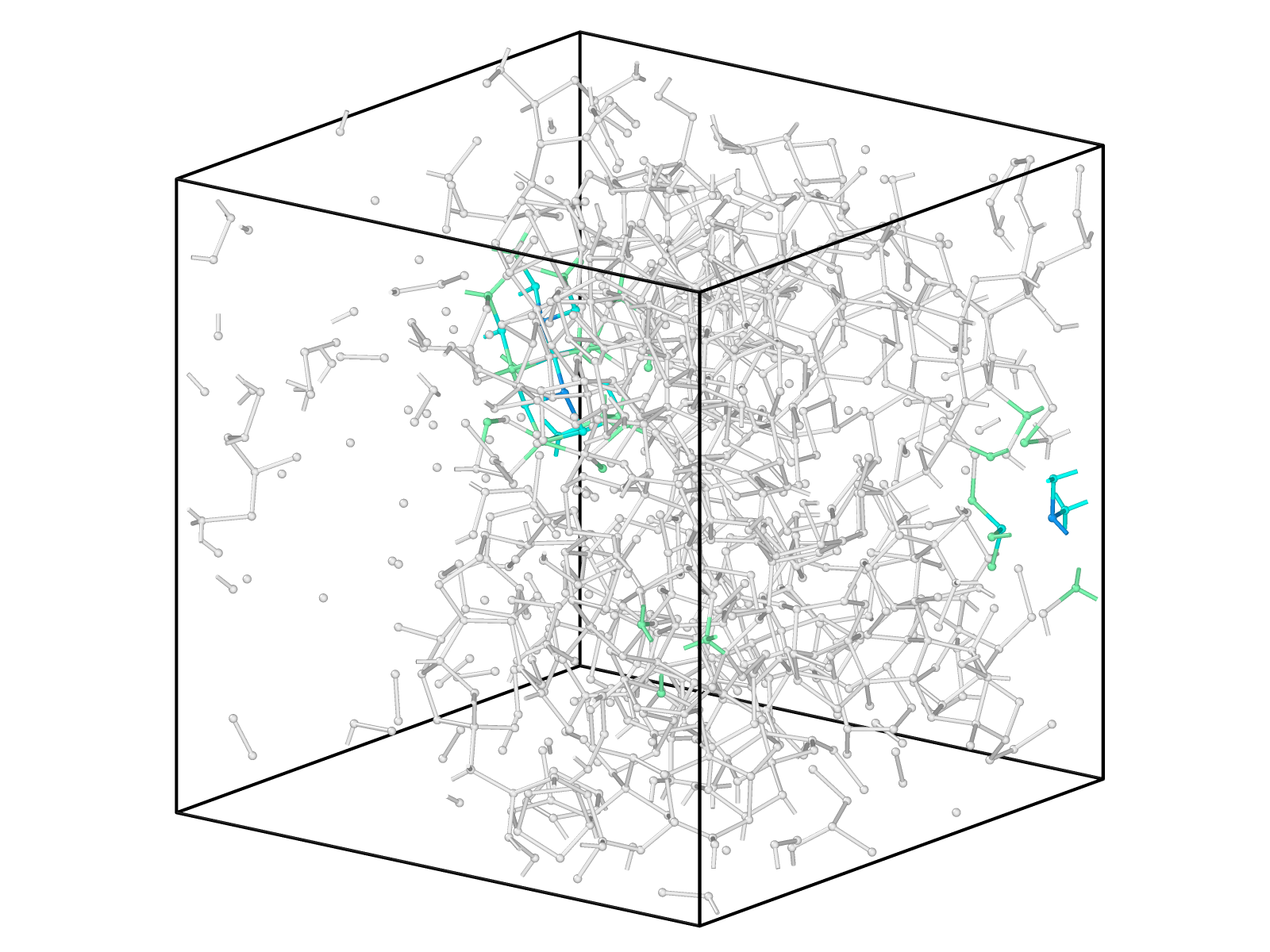}
     }
    \subfloat[\label{fig:box_0097_last}]{
    \includegraphics[width=0.33\textwidth]{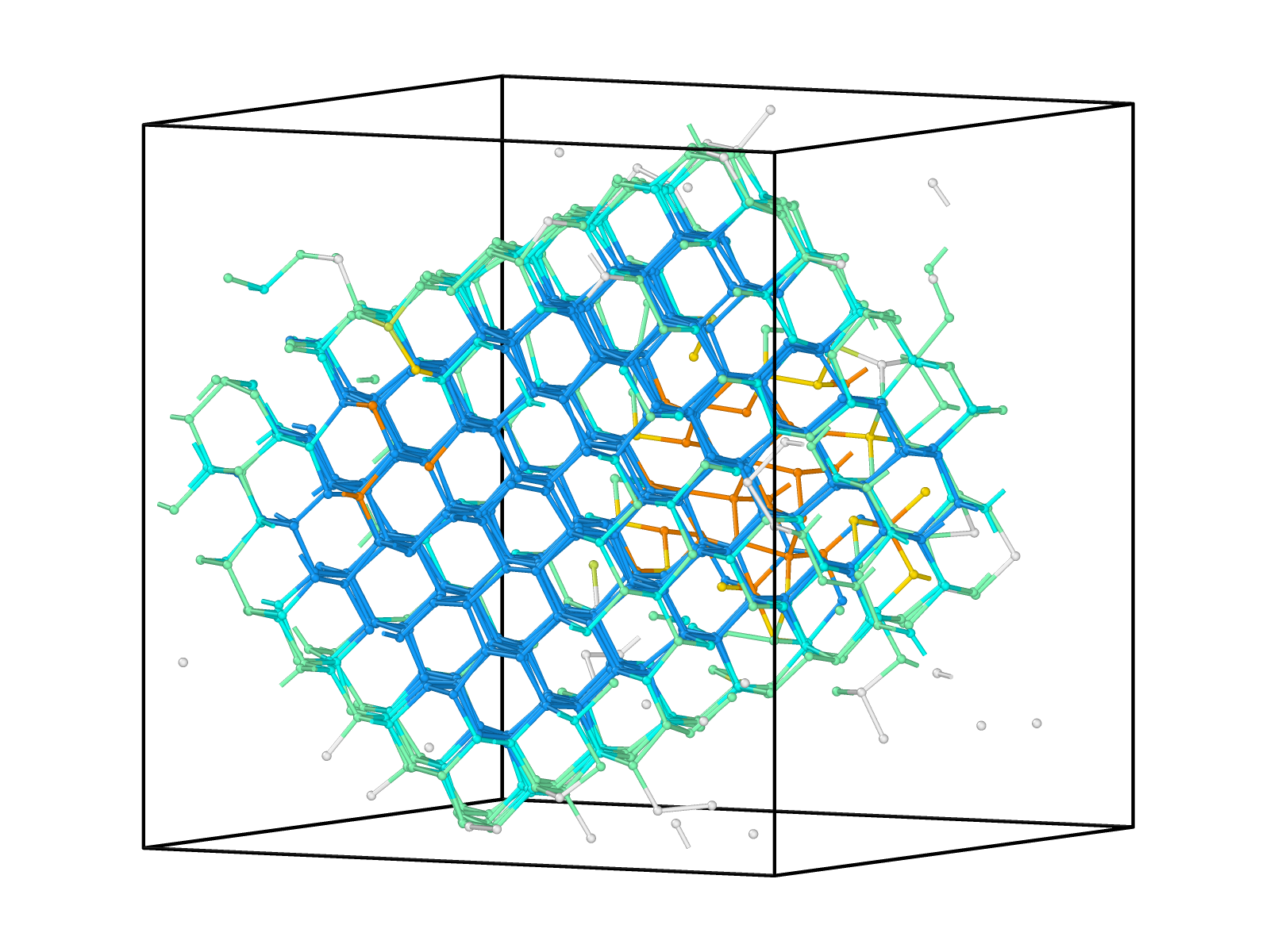}
     }
    \caption{Overview of Monte Carlo simulation with AVB moves in the NVT ensemble run at $T=0.097$, $x=0.5$ and $\rho=0.3$ where $T$ is in unit of $k_{B}/\epsilon$ and $\rho$ is in unit of $1/\sigma^{3}$. a) Potential energy per particle as a function of MC sweeps: the energy drop signals nucleation event. b) Snapshot of a metastable configuration before the energy drop showing that we deal with a two step nucleation since crystalline nucleus arise in the denser liquid phase.  c) Snapshot of the last configuration showing that the N2c8 binary mixture eventually self-assembles into a cubic diamond crystal. Neither patches nor the two patchy particles species are shown while blue/green colors indicate particles belonging to the cubic diamond phase. Snapshots and crystal identification are obtained with the Ovito software~\cite{ovito}.}
    \label{fig_traj_0097}
\end{figure*} 

In Fig.~\ref{fig_traj_0097} we show the evolution of a typical nucleation trajectory at the azeotropic point at $T=0.097$ and $\rho=0.3$. Panel (a) shows the evolution of the energy, displaying a two-step drop: initially the system is prepared in a metastable gas phase, which then forms a percolating liquid phase; in the second step (from $1.9\cdot 10^8$ MC sweeps) the nucleation occurs within the liquid drop and a diamond cubic crystal replaces all the liquid phase, finally coexisting with a low-density gas. Panel (b) shows a snapshot at the end of the first stage, where a liquid-gas coexistence is established in the box. Panel (c) shows a snapshot at the final stage of the trajectory, visually confirming the formation of a large crystal of the diamond cubic phase. The absence of the hexagonal diamond phase and stacking fault defects is a property embedded into the color interactions of the N2c8 mixture, which was designed with this property via the SAT-assembly framework~\cite{rovigatti2022simple}.

\section{Discussion and Conclusions}

Recent experimental advances hold the promise to fully unlock the potential of systems with directional and specific interactions. In particular DNA nanotechnology already allows the fabrication of patchy particles in the form of wireframe DNA origami~\cite{liu2016diamond,zhang20183d,tian2020ordered,chakraborty2022self} and gold nanoparticles with selective patterning~\cite{xiong2020three}.
DNA strands naturally encode sequence complementary and thus allow the assembly of structures with the desired interactions: different interacting patches correspond to complementary single DNA strands as well as the self-interacting ones correspond to palindromic DNA strands.
This class of systems goes under the name of \emph{Empty Liquids}~\cite{russo2021physics}, which have unique properties compared to simple (hard-sphere like) systems, such as the existence of low-density stable liquid phases~\cite{bianchi2006empty}, the ability to crystallize in open crystals~\cite{romano2011crystallization}, and the potential to model the phase bahaviour of biological molecules~\cite{altan2019obtaining,gnan2019patchy,mcmanus2019protein,hvozd2022empty}.

To obtain a complex target structure (such as a photonic crystal) requires either a complex interaction potential carefully optimized to have the target as a free-energy minima, or a mixture of simpler building blocks that make the target structure accessible by reducing its overall symmetry (crystal sites are occupied by different species). Recent advances in DNA-nanotechnology, allowing the design of building blocks with arbitrary interaction rules, are opening the door to this last approach.

Several strategies have been proposed to design the interaction potential in multi-component system. One such strategy is the SAT-assembly framework, which translates the design problem in a Boolean satisfiability problem~\cite{romano2020designing,russo2022sat}. The ability to design interaction matrices to self-assemble arbitrary structures is a necessary but not sufficient condition for observing the nucleation of such structures under experimental conditions. These often involve preparing the system in very dilute conditions, from which the classical nucleation rate is severely suppressed by both diffusional and compositional barriers.

 Regarding the diffusional barrier, probably the most promising strategy to overcome it is \emph{two-step} nucleation, where the crystal nucleation is aided by the formation of dense liquid drops. This nucleation pathway has been the subject of intense research, especially in the context of one-component systems.

So far not much work has been devoted to the second barrier, i.e. the compositional barrier, which controls the rate at which a particle of the correct species enters in contact with the incipient nucleus.
In this work we have presented a first study in this direction. We have examined a two-step nucleation process in a binary mixture of patchy particles designed to assemble into cubic diamond. In particular we have focused on the role of the metastable liquid-gas phase diagram on the nucleation process. Comparing Gibbs ensemble simulations with direct nucleation runs, we have established that nucleating state points coincide with the locus of metastability of the mixture, confirming the relevant role played by the thermodynamic instabilities, generating local environments with a density enhanced in comparison to the average value. We have shown that composition effects in multi-component systems play a big role in the growth of the crystal, and that crystal growth
is enhanced at state points where the emerging liquid phase has a composition close to the 
 stoichiometric composition of the crystal. In the studied system, this corresponds to an azeotropic point at equimolar concentration, $x=0.5$. We  find evidence of critical-like enhanced nucleation along the binary critical line when this intersects the azeotropic line.

To unlock the promises of nanotechnology to self-assemble new materials with desired mechanical, optical and thermal properties, a big role will be played by the search of general principles behind the crystallization of multi-component systems. We believe that two-step nucleation, augmented with azeotropic conditions, will be one of such guiding principles.

\section{Acknowledgments}
We acknowledge the CINECA award NNPROT under the ISCRA initiative, for the availability of high performance computing resources and support.
JR and DEPP acknowledge support from the European Research Council Grant DLV-759187. 
 This result is part of a project that has received funding from the European Research Council (ERC) under the European Union’s Horizon 2020 research and innovation programme (Grant agreement No. 101040035) (to P\v{S})
 

\begin{thebibliography}{85}%
\makeatletter
\providecommand \@ifxundefined [1]{%
 \@ifx{#1\undefined}
}%
\providecommand \@ifnum [1]{%
 \ifnum #1\expandafter \@firstoftwo
 \else \expandafter \@secondoftwo
 \fi
}%
\providecommand \@ifx [1]{%
 \ifx #1\expandafter \@firstoftwo
 \else \expandafter \@secondoftwo
 \fi
}%
\providecommand \natexlab [1]{#1}%
\providecommand \enquote  [1]{``#1''}%
\providecommand \bibnamefont  [1]{#1}%
\providecommand \bibfnamefont [1]{#1}%
\providecommand \citenamefont [1]{#1}%
\providecommand \href@noop [0]{\@secondoftwo}%
\providecommand \href [0]{\begingroup \@sanitize@url \@href}%
\providecommand \@href[1]{\@@startlink{#1}\@@href}%
\providecommand \@@href[1]{\endgroup#1\@@endlink}%
\providecommand \@sanitize@url [0]{\catcode `\\12\catcode `\$12\catcode
  `\&12\catcode `\#12\catcode `\^12\catcode `\_12\catcode `\%12\relax}%
\providecommand \@@startlink[1]{}%
\providecommand \@@endlink[0]{}%
\providecommand \url  [0]{\begingroup\@sanitize@url \@url }%
\providecommand \@url [1]{\endgroup\@href {#1}{\urlprefix }}%
\providecommand \urlprefix  [0]{URL }%
\providecommand \Eprint [0]{\href }%
\providecommand \doibase [0]{https://doi.org/}%
\providecommand \selectlanguage [0]{\@gobble}%
\providecommand \bibinfo  [0]{\@secondoftwo}%
\providecommand \bibfield  [0]{\@secondoftwo}%
\providecommand \translation [1]{[#1]}%
\providecommand \BibitemOpen [0]{}%
\providecommand \bibitemStop [0]{}%
\providecommand \bibitemNoStop [0]{.\EOS\space}%
\providecommand \EOS [0]{\spacefactor3000\relax}%
\providecommand \BibitemShut  [1]{\csname bibitem#1\endcsname}%
\let\auto@bib@innerbib\@empty
\bibitem [{\citenamefont {Whitelam}\ and\ \citenamefont
  {Jack}(2015)}]{whitelam2015statistical}%
  \BibitemOpen
  \bibfield  {author} {\bibinfo {author} {\bibfnamefont {S.}~\bibnamefont
  {Whitelam}}\ and\ \bibinfo {author} {\bibfnamefont {R.~L.}\ \bibnamefont
  {Jack}},\ }\bibfield  {title} {\bibinfo {title} {The statistical mechanics of
  dynamic pathways to self-assembly},\ }\href@noop {} {\bibfield  {journal}
  {\bibinfo  {journal} {Annual review of physical chemistry}\ }\textbf
  {\bibinfo {volume} {66}},\ \bibinfo {pages} {143} (\bibinfo {year}
  {2015})}\BibitemShut {NoStop}%
\bibitem [{\citenamefont {Kumar}\ \emph {et~al.}(2017)\citenamefont {Kumar},
  \citenamefont {Kumaraswamy}, \citenamefont {Prasad}, \citenamefont
  {Bandyopadhyaya}, \citenamefont {Granick}, \citenamefont {Gang},
  \citenamefont {Manoharan}, \citenamefont {Frenkel},\ and\ \citenamefont
  {Kotov}}]{kumar2017nanoparticle}%
  \BibitemOpen
  \bibfield  {author} {\bibinfo {author} {\bibfnamefont {S.~K.}\ \bibnamefont
  {Kumar}}, \bibinfo {author} {\bibfnamefont {G.}~\bibnamefont {Kumaraswamy}},
  \bibinfo {author} {\bibfnamefont {B.~L.}\ \bibnamefont {Prasad}}, \bibinfo
  {author} {\bibfnamefont {R.}~\bibnamefont {Bandyopadhyaya}}, \bibinfo
  {author} {\bibfnamefont {S.}~\bibnamefont {Granick}}, \bibinfo {author}
  {\bibfnamefont {O.}~\bibnamefont {Gang}}, \bibinfo {author} {\bibfnamefont
  {V.~N.}\ \bibnamefont {Manoharan}}, \bibinfo {author} {\bibfnamefont
  {D.}~\bibnamefont {Frenkel}},\ and\ \bibinfo {author} {\bibfnamefont {N.~A.}\
  \bibnamefont {Kotov}},\ }\bibfield  {title} {\bibinfo {title} {Nanoparticle
  assembly: a perspective and some unanswered questions},\ }\href@noop {}
  {\bibfield  {journal} {\bibinfo  {journal} {Current Science}\ ,\ \bibinfo
  {pages} {1635}} (\bibinfo {year} {2017})}\BibitemShut {NoStop}%
\bibitem [{\citenamefont {Seeman}\ and\ \citenamefont
  {Sleiman}(2017)}]{seeman2017dna}%
  \BibitemOpen
  \bibfield  {author} {\bibinfo {author} {\bibfnamefont {N.~C.}\ \bibnamefont
  {Seeman}}\ and\ \bibinfo {author} {\bibfnamefont {H.~F.}\ \bibnamefont
  {Sleiman}},\ }\bibfield  {title} {\bibinfo {title} {Dna nanotechnology},\
  }\href@noop {} {\bibfield  {journal} {\bibinfo  {journal} {Nature Reviews
  Materials}\ }\textbf {\bibinfo {volume} {3}},\ \bibinfo {pages} {1} (\bibinfo
  {year} {2017})}\BibitemShut {NoStop}%
\bibitem [{\citenamefont {Liu}\ \emph {et~al.}(2016)\citenamefont {Liu},
  \citenamefont {Tagawa}, \citenamefont {Xin}, \citenamefont {Wang},
  \citenamefont {Emamy}, \citenamefont {Li}, \citenamefont {Yager},
  \citenamefont {Starr}, \citenamefont {Tkachenko},\ and\ \citenamefont
  {Gang}}]{liu2016diamond}%
  \BibitemOpen
  \bibfield  {author} {\bibinfo {author} {\bibfnamefont {W.}~\bibnamefont
  {Liu}}, \bibinfo {author} {\bibfnamefont {M.}~\bibnamefont {Tagawa}},
  \bibinfo {author} {\bibfnamefont {H.~L.}\ \bibnamefont {Xin}}, \bibinfo
  {author} {\bibfnamefont {T.}~\bibnamefont {Wang}}, \bibinfo {author}
  {\bibfnamefont {H.}~\bibnamefont {Emamy}}, \bibinfo {author} {\bibfnamefont
  {H.}~\bibnamefont {Li}}, \bibinfo {author} {\bibfnamefont {K.~G.}\
  \bibnamefont {Yager}}, \bibinfo {author} {\bibfnamefont {F.~W.}\ \bibnamefont
  {Starr}}, \bibinfo {author} {\bibfnamefont {A.~V.}\ \bibnamefont
  {Tkachenko}},\ and\ \bibinfo {author} {\bibfnamefont {O.}~\bibnamefont
  {Gang}},\ }\bibfield  {title} {\bibinfo {title} {Diamond family of
  nanoparticle superlattices},\ }\href@noop {} {\bibfield  {journal} {\bibinfo
  {journal} {Science}\ }\textbf {\bibinfo {volume} {351}},\ \bibinfo {pages}
  {582} (\bibinfo {year} {2016})}\BibitemShut {NoStop}%
\bibitem [{\citenamefont {Zhang}\ \emph {et~al.}(2018)\citenamefont {Zhang},
  \citenamefont {Hartl}, \citenamefont {Frank}, \citenamefont
  {Heuer-Jungemann}, \citenamefont {Fischer}, \citenamefont {Nickels},
  \citenamefont {Nickel},\ and\ \citenamefont {Liedl}}]{zhang20183d}%
  \BibitemOpen
  \bibfield  {author} {\bibinfo {author} {\bibfnamefont {T.}~\bibnamefont
  {Zhang}}, \bibinfo {author} {\bibfnamefont {C.}~\bibnamefont {Hartl}},
  \bibinfo {author} {\bibfnamefont {K.}~\bibnamefont {Frank}}, \bibinfo
  {author} {\bibfnamefont {A.}~\bibnamefont {Heuer-Jungemann}}, \bibinfo
  {author} {\bibfnamefont {S.}~\bibnamefont {Fischer}}, \bibinfo {author}
  {\bibfnamefont {P.~C.}\ \bibnamefont {Nickels}}, \bibinfo {author}
  {\bibfnamefont {B.}~\bibnamefont {Nickel}},\ and\ \bibinfo {author}
  {\bibfnamefont {T.}~\bibnamefont {Liedl}},\ }\bibfield  {title} {\bibinfo
  {title} {3d dna origami crystals},\ }\href@noop {} {\bibfield  {journal}
  {\bibinfo  {journal} {Advanced Materials}\ }\textbf {\bibinfo {volume}
  {30}},\ \bibinfo {pages} {1800273} (\bibinfo {year} {2018})}\BibitemShut
  {NoStop}%
\bibitem [{\citenamefont {Kashchiev}\ \emph {et~al.}(2005)\citenamefont
  {Kashchiev}, \citenamefont {Vekilov},\ and\ \citenamefont
  {Kolomeisky}}]{kashchiev2005kinetics}%
  \BibitemOpen
  \bibfield  {author} {\bibinfo {author} {\bibfnamefont {D.}~\bibnamefont
  {Kashchiev}}, \bibinfo {author} {\bibfnamefont {P.~G.}\ \bibnamefont
  {Vekilov}},\ and\ \bibinfo {author} {\bibfnamefont {A.~B.}\ \bibnamefont
  {Kolomeisky}},\ }\bibfield  {title} {\bibinfo {title} {Kinetics of two-step
  nucleation of crystals},\ }\href@noop {} {\bibfield  {journal} {\bibinfo
  {journal} {The Journal of chemical physics}\ }\textbf {\bibinfo {volume}
  {122}},\ \bibinfo {pages} {244706} (\bibinfo {year} {2005})}\BibitemShut
  {NoStop}%
\bibitem [{\citenamefont {Erdemir}\ \emph {et~al.}(2009)\citenamefont
  {Erdemir}, \citenamefont {Lee},\ and\ \citenamefont
  {Myerson}}]{erdemir2009nucleation}%
  \BibitemOpen
  \bibfield  {author} {\bibinfo {author} {\bibfnamefont {D.}~\bibnamefont
  {Erdemir}}, \bibinfo {author} {\bibfnamefont {A.~Y.}\ \bibnamefont {Lee}},\
  and\ \bibinfo {author} {\bibfnamefont {A.~S.}\ \bibnamefont {Myerson}},\
  }\bibfield  {title} {\bibinfo {title} {Nucleation of crystals from solution:
  classical and two-step models},\ }\href@noop {} {\bibfield  {journal}
  {\bibinfo  {journal} {Accounts of chemical research}\ }\textbf {\bibinfo
  {volume} {42}},\ \bibinfo {pages} {621} (\bibinfo {year} {2009})}\BibitemShut
  {NoStop}%
\bibitem [{\citenamefont {Vekilov}(2010)}]{vekilov2010two}%
  \BibitemOpen
  \bibfield  {author} {\bibinfo {author} {\bibfnamefont {P.~G.}\ \bibnamefont
  {Vekilov}},\ }\bibfield  {title} {\bibinfo {title} {The two-step mechanism of
  nucleation of crystals in solution},\ }\href@noop {} {\bibfield  {journal}
  {\bibinfo  {journal} {Nanoscale}\ }\textbf {\bibinfo {volume} {2}},\ \bibinfo
  {pages} {2346} (\bibinfo {year} {2010})}\BibitemShut {NoStop}%
\bibitem [{\citenamefont {T{\'o}th}\ \emph {et~al.}(2011)\citenamefont
  {T{\'o}th}, \citenamefont {Pusztai}, \citenamefont {Tegze}, \citenamefont
  {T{\'o}th},\ and\ \citenamefont {Gr{\'a}n{\'a}sy}}]{toth2011amorphous}%
  \BibitemOpen
  \bibfield  {author} {\bibinfo {author} {\bibfnamefont {G.~I.}\ \bibnamefont
  {T{\'o}th}}, \bibinfo {author} {\bibfnamefont {T.}~\bibnamefont {Pusztai}},
  \bibinfo {author} {\bibfnamefont {G.}~\bibnamefont {Tegze}}, \bibinfo
  {author} {\bibfnamefont {G.}~\bibnamefont {T{\'o}th}},\ and\ \bibinfo
  {author} {\bibfnamefont {L.}~\bibnamefont {Gr{\'a}n{\'a}sy}},\ }\bibfield
  {title} {\bibinfo {title} {Amorphous nucleation precursor in highly
  nonequilibrium fluids},\ }\href@noop {} {\bibfield  {journal} {\bibinfo
  {journal} {Phys. Rev. Lett.}\ }\textbf {\bibinfo {volume} {107}},\ \bibinfo
  {pages} {175702} (\bibinfo {year} {2011})}\BibitemShut {NoStop}%
\bibitem [{\citenamefont {Sear}(2012)}]{sear2012}%
  \BibitemOpen
  \bibfield  {author} {\bibinfo {author} {\bibfnamefont {R.~P.}\ \bibnamefont
  {Sear}},\ }\bibfield  {title} {\bibinfo {title} {The non-classical nucleation
  of crystals: microscopic mechanisms and applications to molecular crystals,
  ice and calcium carbonate},\ }\href@noop {} {\bibfield  {journal} {\bibinfo
  {journal} {International Materials Reviews}\ }\textbf {\bibinfo {volume}
  {57}},\ \bibinfo {pages} {328} (\bibinfo {year} {2012})}\BibitemShut
  {NoStop}%
\bibitem [{\citenamefont {Haxton}\ \emph {et~al.}(2015)\citenamefont {Haxton},
  \citenamefont {Hedges},\ and\ \citenamefont
  {Whitelam}}]{haxton2015crystallization}%
  \BibitemOpen
  \bibfield  {author} {\bibinfo {author} {\bibfnamefont {T.~K.}\ \bibnamefont
  {Haxton}}, \bibinfo {author} {\bibfnamefont {L.~O.}\ \bibnamefont {Hedges}},\
  and\ \bibinfo {author} {\bibfnamefont {S.}~\bibnamefont {Whitelam}},\
  }\bibfield  {title} {\bibinfo {title} {Crystallization and arrest mechanisms
  of model colloids},\ }\href@noop {} {\bibfield  {journal} {\bibinfo
  {journal} {Soft matter}\ }\textbf {\bibinfo {volume} {11}},\ \bibinfo {pages}
  {9307} (\bibinfo {year} {2015})}\BibitemShut {NoStop}%
\bibitem [{\citenamefont {Russo}\ and\ \citenamefont
  {Tanaka}(2016)}]{russo2016nonclassical}%
  \BibitemOpen
  \bibfield  {author} {\bibinfo {author} {\bibfnamefont {J.}~\bibnamefont
  {Russo}}\ and\ \bibinfo {author} {\bibfnamefont {H.}~\bibnamefont {Tanaka}},\
  }\bibfield  {title} {\bibinfo {title} {Nonclassical pathways of
  crystallization in colloidal systems},\ }\href@noop {} {\bibfield  {journal}
  {\bibinfo  {journal} {MRS Bulletin}\ }\textbf {\bibinfo {volume} {41}},\
  \bibinfo {pages} {369} (\bibinfo {year} {2016})}\BibitemShut {NoStop}%
\bibitem [{\citenamefont {Sosso}\ \emph {et~al.}(2016)\citenamefont {Sosso},
  \citenamefont {Chen}, \citenamefont {Cox}, \citenamefont {Fitzner},
  \citenamefont {Pedevilla}, \citenamefont {Zen},\ and\ \citenamefont
  {Michaelides}}]{sosso2016crystal}%
  \BibitemOpen
  \bibfield  {author} {\bibinfo {author} {\bibfnamefont {G.~C.}\ \bibnamefont
  {Sosso}}, \bibinfo {author} {\bibfnamefont {J.}~\bibnamefont {Chen}},
  \bibinfo {author} {\bibfnamefont {S.~J.}\ \bibnamefont {Cox}}, \bibinfo
  {author} {\bibfnamefont {M.}~\bibnamefont {Fitzner}}, \bibinfo {author}
  {\bibfnamefont {P.}~\bibnamefont {Pedevilla}}, \bibinfo {author}
  {\bibfnamefont {A.}~\bibnamefont {Zen}},\ and\ \bibinfo {author}
  {\bibfnamefont {A.}~\bibnamefont {Michaelides}},\ }\bibfield  {title}
  {\bibinfo {title} {Crystal nucleation in liquids: Open questions and future
  challenges in molecular dynamics simulations},\ }\href@noop {} {\bibfield
  {journal} {\bibinfo  {journal} {Chemical reviews}\ }\textbf {\bibinfo
  {volume} {116}},\ \bibinfo {pages} {7078} (\bibinfo {year}
  {2016})}\BibitemShut {NoStop}%
\bibitem [{\citenamefont {Lutsko}(2019)}]{lutsko2019crystals}%
  \BibitemOpen
  \bibfield  {author} {\bibinfo {author} {\bibfnamefont {J.~F.}\ \bibnamefont
  {Lutsko}},\ }\bibfield  {title} {\bibinfo {title} {How crystals form: A
  theory of nucleation pathways},\ }\href@noop {} {\bibfield  {journal}
  {\bibinfo  {journal} {Sci. Adv.}\ }\textbf {\bibinfo {volume} {5}},\ \bibinfo
  {pages} {eaav7399} (\bibinfo {year} {2019})}\BibitemShut {NoStop}%
\bibitem [{\citenamefont {James}\ \emph {et~al.}(2019)\citenamefont {James},
  \citenamefont {Beairsto}, \citenamefont {Hartt}, \citenamefont {Zavalov},
  \citenamefont {Saika-Voivod}, \citenamefont {Bowles},\ and\ \citenamefont
  {Poole}}]{james2019phase}%
  \BibitemOpen
  \bibfield  {author} {\bibinfo {author} {\bibfnamefont {D.}~\bibnamefont
  {James}}, \bibinfo {author} {\bibfnamefont {S.}~\bibnamefont {Beairsto}},
  \bibinfo {author} {\bibfnamefont {C.}~\bibnamefont {Hartt}}, \bibinfo
  {author} {\bibfnamefont {O.}~\bibnamefont {Zavalov}}, \bibinfo {author}
  {\bibfnamefont {I.}~\bibnamefont {Saika-Voivod}}, \bibinfo {author}
  {\bibfnamefont {R.~K.}\ \bibnamefont {Bowles}},\ and\ \bibinfo {author}
  {\bibfnamefont {P.~H.}\ \bibnamefont {Poole}},\ }\bibfield  {title} {\bibinfo
  {title} {Phase transitions in fluctuations and their role in two-step
  nucleation},\ }\href@noop {} {\bibfield  {journal} {\bibinfo  {journal} {J.
  Chem. Phys.}\ }\textbf {\bibinfo {volume} {150}},\ \bibinfo {pages} {074501}
  (\bibinfo {year} {2019})}\BibitemShut {NoStop}%
\bibitem [{\citenamefont {Desgranges}\ and\ \citenamefont
  {Delhommelle}(2019)}]{desgranges2019can}%
  \BibitemOpen
  \bibfield  {author} {\bibinfo {author} {\bibfnamefont {C.}~\bibnamefont
  {Desgranges}}\ and\ \bibinfo {author} {\bibfnamefont {J.}~\bibnamefont
  {Delhommelle}},\ }\bibfield  {title} {\bibinfo {title} {Can ordered
  precursors promote the nucleation of solid solutions?},\ }\href@noop {}
  {\bibfield  {journal} {\bibinfo  {journal} {Phys. Rev. Lett.}\ }\textbf
  {\bibinfo {volume} {123}},\ \bibinfo {pages} {195701} (\bibinfo {year}
  {2019})}\BibitemShut {NoStop}%
\bibitem [{\citenamefont {Kashchiev}(2020)}]{kashchiev2020classical}%
  \BibitemOpen
  \bibfield  {author} {\bibinfo {author} {\bibfnamefont {D.}~\bibnamefont
  {Kashchiev}},\ }\bibfield  {title} {\bibinfo {title} {Classical nucleation
  theory approach to two-step nucleation of crystals},\ }\href@noop {}
  {\bibfield  {journal} {\bibinfo  {journal} {Journal of Crystal Growth}\
  }\textbf {\bibinfo {volume} {530}},\ \bibinfo {pages} {125300} (\bibinfo
  {year} {2020})}\BibitemShut {NoStop}%
\bibitem [{\citenamefont {Lee}\ \emph {et~al.}(2019)\citenamefont {Lee},
  \citenamefont {Teich}, \citenamefont {Engel},\ and\ \citenamefont
  {Glotzer}}]{lee2019entropic}%
  \BibitemOpen
  \bibfield  {author} {\bibinfo {author} {\bibfnamefont {S.}~\bibnamefont
  {Lee}}, \bibinfo {author} {\bibfnamefont {E.~G.}\ \bibnamefont {Teich}},
  \bibinfo {author} {\bibfnamefont {M.}~\bibnamefont {Engel}},\ and\ \bibinfo
  {author} {\bibfnamefont {S.~C.}\ \bibnamefont {Glotzer}},\ }\bibfield
  {title} {\bibinfo {title} {Entropic colloidal crystallization pathways via
  fluid--fluid transitions and multidimensional prenucleation motifs},\
  }\href@noop {} {\bibfield  {journal} {\bibinfo  {journal} {Proceedings of the
  National Academy of Sciences}\ }\textbf {\bibinfo {volume} {116}},\ \bibinfo
  {pages} {14843} (\bibinfo {year} {2019})}\BibitemShut {NoStop}%
\bibitem [{\citenamefont {ten Wolde}\ and\ \citenamefont
  {Frenkel}(1997)}]{ten1997enhancement}%
  \BibitemOpen
  \bibfield  {author} {\bibinfo {author} {\bibfnamefont {P.~R.}\ \bibnamefont
  {ten Wolde}}\ and\ \bibinfo {author} {\bibfnamefont {D.}~\bibnamefont
  {Frenkel}},\ }\bibfield  {title} {\bibinfo {title} {Enhancement of protein
  crystal nucleation by critical density fluctuations},\ }\href@noop {}
  {\bibfield  {journal} {\bibinfo  {journal} {Science}\ }\textbf {\bibinfo
  {volume} {277}},\ \bibinfo {pages} {1975} (\bibinfo {year}
  {1997})}\BibitemShut {NoStop}%
\bibitem [{\citenamefont {Tan}\ \emph {et~al.}(2014)\citenamefont {Tan},
  \citenamefont {Xu},\ and\ \citenamefont {Xu}}]{tan2014}%
  \BibitemOpen
  \bibfield  {author} {\bibinfo {author} {\bibfnamefont {P.}~\bibnamefont
  {Tan}}, \bibinfo {author} {\bibfnamefont {N.}~\bibnamefont {Xu}},\ and\
  \bibinfo {author} {\bibfnamefont {L.}~\bibnamefont {Xu}},\ }\bibfield
  {title} {\bibinfo {title} {Visualizing kinetic pathways of homogeneous
  nucleation in colloidal crystallization},\ }\href@noop {} {\bibfield
  {journal} {\bibinfo  {journal} {Nature Physics}\ }\textbf {\bibinfo {volume}
  {10}},\ \bibinfo {pages} {73} (\bibinfo {year} {2014})}\BibitemShut {NoStop}%
\bibitem [{\citenamefont {Jiang}\ \emph {et~al.}(2019)\citenamefont {Jiang},
  \citenamefont {Debenedetti},\ and\ \citenamefont
  {Panagiotopoulos}}]{jiang2019nucleation}%
  \BibitemOpen
  \bibfield  {author} {\bibinfo {author} {\bibfnamefont {H.}~\bibnamefont
  {Jiang}}, \bibinfo {author} {\bibfnamefont {P.~G.}\ \bibnamefont
  {Debenedetti}},\ and\ \bibinfo {author} {\bibfnamefont {A.~Z.}\ \bibnamefont
  {Panagiotopoulos}},\ }\bibfield  {title} {\bibinfo {title} {Nucleation in
  aqueous nacl solutions shifts from 1-step to 2-step mechanism on crossing the
  spinodal},\ }\href@noop {} {\bibfield  {journal} {\bibinfo  {journal} {The
  Journal of chemical physics}\ }\textbf {\bibinfo {volume} {150}},\ \bibinfo
  {pages} {124502} (\bibinfo {year} {2019})}\BibitemShut {NoStop}%
\bibitem [{\citenamefont {Gebauer}\ \emph {et~al.}(2008)\citenamefont
  {Gebauer}, \citenamefont {V{\"o}lkel},\ and\ \citenamefont
  {C{\"o}lfen}}]{gebauer2008}%
  \BibitemOpen
  \bibfield  {author} {\bibinfo {author} {\bibfnamefont {D.}~\bibnamefont
  {Gebauer}}, \bibinfo {author} {\bibfnamefont {A.}~\bibnamefont
  {V{\"o}lkel}},\ and\ \bibinfo {author} {\bibfnamefont {H.}~\bibnamefont
  {C{\"o}lfen}},\ }\bibfield  {title} {\bibinfo {title} {Stable prenucleation
  calcium carbonate clusters},\ }\href@noop {} {\bibfield  {journal} {\bibinfo
  {journal} {Science}\ }\textbf {\bibinfo {volume} {322}},\ \bibinfo {pages}
  {1819} (\bibinfo {year} {2008})}\BibitemShut {NoStop}%
\bibitem [{\citenamefont {Pouget}\ \emph {et~al.}(2009)\citenamefont {Pouget},
  \citenamefont {Bomans}, \citenamefont {Goos}, \citenamefont {Frederik},
  \citenamefont {de~With},\ and\ \citenamefont {Sommerdijk}}]{pouget2009}%
  \BibitemOpen
  \bibfield  {author} {\bibinfo {author} {\bibfnamefont {E.~M.}\ \bibnamefont
  {Pouget}}, \bibinfo {author} {\bibfnamefont {P.~H.~H.}\ \bibnamefont
  {Bomans}}, \bibinfo {author} {\bibfnamefont {J.~A. C.~M.}\ \bibnamefont
  {Goos}}, \bibinfo {author} {\bibfnamefont {P.~M.}\ \bibnamefont {Frederik}},
  \bibinfo {author} {\bibfnamefont {G.}~\bibnamefont {de~With}},\ and\ \bibinfo
  {author} {\bibfnamefont {N.~A. J.~M.}\ \bibnamefont {Sommerdijk}},\
  }\bibfield  {title} {\bibinfo {title} {The initial stages of
  template-controlled caco3 formation revealed by cryo-tem},\ }\href@noop {}
  {\bibfield  {journal} {\bibinfo  {journal} {Science}\ }\textbf {\bibinfo
  {volume} {323}},\ \bibinfo {pages} {1455} (\bibinfo {year}
  {2009})}\BibitemShut {NoStop}%
\bibitem [{\citenamefont {Fusco}\ and\ \citenamefont
  {Charbonneau}(2013)}]{fusco2013crystallization}%
  \BibitemOpen
  \bibfield  {author} {\bibinfo {author} {\bibfnamefont {D.}~\bibnamefont
  {Fusco}}\ and\ \bibinfo {author} {\bibfnamefont {P.}~\bibnamefont
  {Charbonneau}},\ }\bibfield  {title} {\bibinfo {title} {Crystallization of
  asymmetric patchy models for globular proteins in solution},\ }\href@noop {}
  {\bibfield  {journal} {\bibinfo  {journal} {Physical Review E}\ }\textbf
  {\bibinfo {volume} {88}},\ \bibinfo {pages} {012721} (\bibinfo {year}
  {2013})}\BibitemShut {NoStop}%
\bibitem [{\citenamefont {James}\ \emph {et~al.}(2015)\citenamefont {James},
  \citenamefont {Quinn},\ and\ \citenamefont {McManus}}]{james2015self}%
  \BibitemOpen
  \bibfield  {author} {\bibinfo {author} {\bibfnamefont {S.}~\bibnamefont
  {James}}, \bibinfo {author} {\bibfnamefont {M.~K.}\ \bibnamefont {Quinn}},\
  and\ \bibinfo {author} {\bibfnamefont {J.~J.}\ \bibnamefont {McManus}},\
  }\bibfield  {title} {\bibinfo {title} {The self assembly of proteins; probing
  patchy protein interactions},\ }\href@noop {} {\bibfield  {journal} {\bibinfo
   {journal} {Physical Chemistry Chemical Physics}\ }\textbf {\bibinfo {volume}
  {17}},\ \bibinfo {pages} {5413} (\bibinfo {year} {2015})}\BibitemShut
  {NoStop}%
\bibitem [{\citenamefont {McManus}\ \emph {et~al.}(2016)\citenamefont
  {McManus}, \citenamefont {Charbonneau}, \citenamefont {Zaccarelli},\ and\
  \citenamefont {Asherie}}]{mcmanus2016physics}%
  \BibitemOpen
  \bibfield  {author} {\bibinfo {author} {\bibfnamefont {J.~J.}\ \bibnamefont
  {McManus}}, \bibinfo {author} {\bibfnamefont {P.}~\bibnamefont
  {Charbonneau}}, \bibinfo {author} {\bibfnamefont {E.}~\bibnamefont
  {Zaccarelli}},\ and\ \bibinfo {author} {\bibfnamefont {N.}~\bibnamefont
  {Asherie}},\ }\bibfield  {title} {\bibinfo {title} {The physics of protein
  self-assembly},\ }\href@noop {} {\bibfield  {journal} {\bibinfo  {journal}
  {Current opinion in colloid \& interface science}\ }\textbf {\bibinfo
  {volume} {22}},\ \bibinfo {pages} {73} (\bibinfo {year} {2016})}\BibitemShut
  {NoStop}%
\bibitem [{\citenamefont {Fusco}\ and\ \citenamefont
  {Charbonneau}(2016)}]{fusco2016soft}%
  \BibitemOpen
  \bibfield  {author} {\bibinfo {author} {\bibfnamefont {D.}~\bibnamefont
  {Fusco}}\ and\ \bibinfo {author} {\bibfnamefont {P.}~\bibnamefont
  {Charbonneau}},\ }\bibfield  {title} {\bibinfo {title} {Soft matter
  perspective on protein crystal assembly},\ }\href@noop {} {\bibfield
  {journal} {\bibinfo  {journal} {Colloids and Surfaces B: Biointerfaces}\
  }\textbf {\bibinfo {volume} {137}},\ \bibinfo {pages} {22} (\bibinfo {year}
  {2016})}\BibitemShut {NoStop}%
\bibitem [{\citenamefont {Jee}\ \emph {et~al.}(2016)\citenamefont {Jee},
  \citenamefont {Lou}, \citenamefont {Jang}, \citenamefont {Nagamanasa},\ and\
  \citenamefont {Granick}}]{jee2016nanoparticle}%
  \BibitemOpen
  \bibfield  {author} {\bibinfo {author} {\bibfnamefont {A.-Y.}\ \bibnamefont
  {Jee}}, \bibinfo {author} {\bibfnamefont {K.}~\bibnamefont {Lou}}, \bibinfo
  {author} {\bibfnamefont {H.-S.}\ \bibnamefont {Jang}}, \bibinfo {author}
  {\bibfnamefont {K.~H.}\ \bibnamefont {Nagamanasa}},\ and\ \bibinfo {author}
  {\bibfnamefont {S.}~\bibnamefont {Granick}},\ }\bibfield  {title} {\bibinfo
  {title} {Nanoparticle puzzles and research opportunities that go beyond state
  of the art},\ }\href@noop {} {\bibfield  {journal} {\bibinfo  {journal}
  {Faraday discussions}\ }\textbf {\bibinfo {volume} {186}},\ \bibinfo {pages}
  {11} (\bibinfo {year} {2016})}\BibitemShut {NoStop}%
\bibitem [{\citenamefont {Dijkstra}\ and\ \citenamefont
  {Luijten}(2021)}]{dijkstra2021predictive}%
  \BibitemOpen
  \bibfield  {author} {\bibinfo {author} {\bibfnamefont {M.}~\bibnamefont
  {Dijkstra}}\ and\ \bibinfo {author} {\bibfnamefont {E.}~\bibnamefont
  {Luijten}},\ }\bibfield  {title} {\bibinfo {title} {From predictive modelling
  to machine learning and reverse engineering of colloidal self-assembly},\
  }\href@noop {} {\bibfield  {journal} {\bibinfo  {journal} {Nature Materials}\
  }\textbf {\bibinfo {volume} {20}},\ \bibinfo {pages} {762} (\bibinfo {year}
  {2021})}\BibitemShut {NoStop}%
\bibitem [{\citenamefont {Whitelam}\ and\ \citenamefont
  {Tamblyn}(2020)}]{whitelam2020learning}%
  \BibitemOpen
  \bibfield  {author} {\bibinfo {author} {\bibfnamefont {S.}~\bibnamefont
  {Whitelam}}\ and\ \bibinfo {author} {\bibfnamefont {I.}~\bibnamefont
  {Tamblyn}},\ }\bibfield  {title} {\bibinfo {title} {Learning to grow: Control
  of material self-assembly using evolutionary reinforcement learning},\
  }\href@noop {} {\bibfield  {journal} {\bibinfo  {journal} {Physical Review
  E}\ }\textbf {\bibinfo {volume} {101}},\ \bibinfo {pages} {052604} (\bibinfo
  {year} {2020})}\BibitemShut {NoStop}%
\bibitem [{\citenamefont {Bupathy}\ \emph {et~al.}(2022)\citenamefont
  {Bupathy}, \citenamefont {Frenkel},\ and\ \citenamefont
  {Sastry}}]{bupathy2022temperature}%
  \BibitemOpen
  \bibfield  {author} {\bibinfo {author} {\bibfnamefont {A.}~\bibnamefont
  {Bupathy}}, \bibinfo {author} {\bibfnamefont {D.}~\bibnamefont {Frenkel}},\
  and\ \bibinfo {author} {\bibfnamefont {S.}~\bibnamefont {Sastry}},\
  }\bibfield  {title} {\bibinfo {title} {Temperature protocols to guide
  selective self-assembly of competing structures},\ }\href@noop {} {\bibfield
  {journal} {\bibinfo  {journal} {Proceedings of the National Academy of
  Sciences}\ }\textbf {\bibinfo {volume} {119}},\ \bibinfo {pages}
  {e2119315119} (\bibinfo {year} {2022})}\BibitemShut {NoStop}%
\bibitem [{\citenamefont {Whitelam}\ and\ \citenamefont
  {Tamblyn}(2021)}]{whitelam2021neuroevolutionary}%
  \BibitemOpen
  \bibfield  {author} {\bibinfo {author} {\bibfnamefont {S.}~\bibnamefont
  {Whitelam}}\ and\ \bibinfo {author} {\bibfnamefont {I.}~\bibnamefont
  {Tamblyn}},\ }\bibfield  {title} {\bibinfo {title} {Neuroevolutionary
  learning of particles and protocols for self-assembly},\ }\href@noop {}
  {\bibfield  {journal} {\bibinfo  {journal} {Physical review letters}\
  }\textbf {\bibinfo {volume} {127}},\ \bibinfo {pages} {018003} (\bibinfo
  {year} {2021})}\BibitemShut {NoStop}%
\bibitem [{\citenamefont {Rechtsman}\ \emph {et~al.}(2005)\citenamefont
  {Rechtsman}, \citenamefont {Stillinger},\ and\ \citenamefont
  {Torquato}}]{rechtsman2005optimized}%
  \BibitemOpen
  \bibfield  {author} {\bibinfo {author} {\bibfnamefont {M.~C.}\ \bibnamefont
  {Rechtsman}}, \bibinfo {author} {\bibfnamefont {F.~H.}\ \bibnamefont
  {Stillinger}},\ and\ \bibinfo {author} {\bibfnamefont {S.}~\bibnamefont
  {Torquato}},\ }\bibfield  {title} {\bibinfo {title} {Optimized interactions
  for targeted self-assembly: application to a honeycomb lattice},\ }\href@noop
  {} {\bibfield  {journal} {\bibinfo  {journal} {Physical review letters}\
  }\textbf {\bibinfo {volume} {95}},\ \bibinfo {pages} {228301} (\bibinfo
  {year} {2005})}\BibitemShut {NoStop}%
\bibitem [{\citenamefont {Marcotte}\ \emph {et~al.}(2011)\citenamefont
  {Marcotte}, \citenamefont {Stillinger},\ and\ \citenamefont
  {Torquato}}]{marcotte2011optimized}%
  \BibitemOpen
  \bibfield  {author} {\bibinfo {author} {\bibfnamefont {E.}~\bibnamefont
  {Marcotte}}, \bibinfo {author} {\bibfnamefont {F.~H.}\ \bibnamefont
  {Stillinger}},\ and\ \bibinfo {author} {\bibfnamefont {S.}~\bibnamefont
  {Torquato}},\ }\bibfield  {title} {\bibinfo {title} {Optimized monotonic
  convex pair potentials stabilize low-coordinated crystals},\ }\href@noop {}
  {\bibfield  {journal} {\bibinfo  {journal} {Soft Matter}\ }\textbf {\bibinfo
  {volume} {7}},\ \bibinfo {pages} {2332} (\bibinfo {year} {2011})}\BibitemShut
  {NoStop}%
\bibitem [{\citenamefont {Marcotte}\ \emph {et~al.}(2013)\citenamefont
  {Marcotte}, \citenamefont {Stillinger},\ and\ \citenamefont
  {Torquato}}]{marcotte2013designeddiamond}%
  \BibitemOpen
  \bibfield  {author} {\bibinfo {author} {\bibfnamefont {E.}~\bibnamefont
  {Marcotte}}, \bibinfo {author} {\bibfnamefont {F.~H.}\ \bibnamefont
  {Stillinger}},\ and\ \bibinfo {author} {\bibfnamefont {S.}~\bibnamefont
  {Torquato}},\ }\bibfield  {title} {\bibinfo {title} {Communication: Designed
  diamond ground state via optimized isotropic monotonic pair potentials},\
  }\href {https://doi.org/10.1063/1.4790634} {\bibfield  {journal} {\bibinfo
  {journal} {The Journal of Chemical Physics}\ }\textbf {\bibinfo {volume}
  {138}},\ \bibinfo {pages} {061101} (\bibinfo {year} {2013})}\BibitemShut
  {NoStop}%
\bibitem [{\citenamefont {Zhang}\ \emph {et~al.}(2013)\citenamefont {Zhang},
  \citenamefont {Stillinger},\ and\ \citenamefont
  {Torquato}}]{zhang2013probing}%
  \BibitemOpen
  \bibfield  {author} {\bibinfo {author} {\bibfnamefont {G.}~\bibnamefont
  {Zhang}}, \bibinfo {author} {\bibfnamefont {F.}~\bibnamefont {Stillinger}},\
  and\ \bibinfo {author} {\bibfnamefont {S.}~\bibnamefont {Torquato}},\
  }\bibfield  {title} {\bibinfo {title} {Probing the limitations of isotropic
  pair potentials to produce ground-state structural extremes via inverse
  statistical mechanics},\ }\href@noop {} {\bibfield  {journal} {\bibinfo
  {journal} {Physical Review E}\ }\textbf {\bibinfo {volume} {88}},\ \bibinfo
  {pages} {042309} (\bibinfo {year} {2013})}\BibitemShut {NoStop}%
\bibitem [{\citenamefont {Miskin}\ \emph {et~al.}(2016)\citenamefont {Miskin},
  \citenamefont {Khaira}, \citenamefont {de~Pablo},\ and\ \citenamefont
  {Jaeger}}]{miskin2016turning}%
  \BibitemOpen
  \bibfield  {author} {\bibinfo {author} {\bibfnamefont {M.~Z.}\ \bibnamefont
  {Miskin}}, \bibinfo {author} {\bibfnamefont {G.}~\bibnamefont {Khaira}},
  \bibinfo {author} {\bibfnamefont {J.~J.}\ \bibnamefont {de~Pablo}},\ and\
  \bibinfo {author} {\bibfnamefont {H.~M.}\ \bibnamefont {Jaeger}},\ }\bibfield
   {title} {\bibinfo {title} {Turning statistical physics models into materials
  design engines},\ }\href@noop {} {\bibfield  {journal} {\bibinfo  {journal}
  {Proceedings of the National Academy of Sciences}\ }\textbf {\bibinfo
  {volume} {113}},\ \bibinfo {pages} {34} (\bibinfo {year} {2016})}\BibitemShut
  {NoStop}%
\bibitem [{\citenamefont {Lindquist}\ \emph {et~al.}(2016)\citenamefont
  {Lindquist}, \citenamefont {Jadrich},\ and\ \citenamefont
  {Truskett}}]{lindquist2016communication}%
  \BibitemOpen
  \bibfield  {author} {\bibinfo {author} {\bibfnamefont {B.~A.}\ \bibnamefont
  {Lindquist}}, \bibinfo {author} {\bibfnamefont {R.~B.}\ \bibnamefont
  {Jadrich}},\ and\ \bibinfo {author} {\bibfnamefont {T.~M.}\ \bibnamefont
  {Truskett}},\ }\bibfield  {title} {\bibinfo {title} {Communication: Inverse
  design for self-assembly via on-the-fly optimization},\ }\href@noop {}
  {\bibfield  {journal} {\bibinfo  {journal} {The Journal of Chemical Physics}\
  }\textbf {\bibinfo {volume} {145}},\ \bibinfo {pages} {111101} (\bibinfo
  {year} {2016})}\BibitemShut {NoStop}%
\bibitem [{\citenamefont {Chen}\ \emph {et~al.}(2018)\citenamefont {Chen},
  \citenamefont {Zhang},\ and\ \citenamefont {Torquato}}]{chen2018inverse}%
  \BibitemOpen
  \bibfield  {author} {\bibinfo {author} {\bibfnamefont {D.}~\bibnamefont
  {Chen}}, \bibinfo {author} {\bibfnamefont {G.}~\bibnamefont {Zhang}},\ and\
  \bibinfo {author} {\bibfnamefont {S.}~\bibnamefont {Torquato}},\ }\bibfield
  {title} {\bibinfo {title} {Inverse design of colloidal crystals via optimized
  patchy interactions},\ }\href@noop {} {\bibfield  {journal} {\bibinfo
  {journal} {The Journal of Physical Chemistry B}\ }\textbf {\bibinfo {volume}
  {122}},\ \bibinfo {pages} {8462} (\bibinfo {year} {2018})}\BibitemShut
  {NoStop}%
\bibitem [{\citenamefont {Kumar}\ \emph {et~al.}(2019)\citenamefont {Kumar},
  \citenamefont {Coli}, \citenamefont {Dijkstra},\ and\ \citenamefont
  {Sastry}}]{kumar2019inverse}%
  \BibitemOpen
  \bibfield  {author} {\bibinfo {author} {\bibfnamefont {R.}~\bibnamefont
  {Kumar}}, \bibinfo {author} {\bibfnamefont {G.~M.}\ \bibnamefont {Coli}},
  \bibinfo {author} {\bibfnamefont {M.}~\bibnamefont {Dijkstra}},\ and\
  \bibinfo {author} {\bibfnamefont {S.}~\bibnamefont {Sastry}},\ }\bibfield
  {title} {\bibinfo {title} {Inverse design of charged colloidal particle
  interactions for self assembly into specified crystal structures},\
  }\href@noop {} {\bibfield  {journal} {\bibinfo  {journal} {The Journal of
  chemical physics}\ }\textbf {\bibinfo {volume} {151}},\ \bibinfo {pages}
  {084109} (\bibinfo {year} {2019})}\BibitemShut {NoStop}%
\bibitem [{\citenamefont {Ducrot}\ \emph {et~al.}(2017)\citenamefont {Ducrot},
  \citenamefont {He}, \citenamefont {Yi},\ and\ \citenamefont
  {Pine}}]{ducrot2017colloidal}%
  \BibitemOpen
  \bibfield  {author} {\bibinfo {author} {\bibfnamefont {{\'E}.}~\bibnamefont
  {Ducrot}}, \bibinfo {author} {\bibfnamefont {M.}~\bibnamefont {He}}, \bibinfo
  {author} {\bibfnamefont {G.-R.}\ \bibnamefont {Yi}},\ and\ \bibinfo {author}
  {\bibfnamefont {D.~J.}\ \bibnamefont {Pine}},\ }\bibfield  {title} {\bibinfo
  {title} {Colloidal alloys with preassembled clusters and spheres},\
  }\href@noop {} {\bibfield  {journal} {\bibinfo  {journal} {Nature materials}\
  }\textbf {\bibinfo {volume} {16}},\ \bibinfo {pages} {652} (\bibinfo {year}
  {2017})}\BibitemShut {NoStop}%
\bibitem [{\citenamefont {Nelson}(2002)}]{nelson2002toward}%
  \BibitemOpen
  \bibfield  {author} {\bibinfo {author} {\bibfnamefont {D.~R.}\ \bibnamefont
  {Nelson}},\ }\bibfield  {title} {\bibinfo {title} {Toward a tetravalent
  chemistry of colloids},\ }\href@noop {} {\bibfield  {journal} {\bibinfo
  {journal} {Nano Letters}\ }\textbf {\bibinfo {volume} {2}},\ \bibinfo {pages}
  {1125} (\bibinfo {year} {2002})}\BibitemShut {NoStop}%
\bibitem [{\citenamefont {Manoharan}\ \emph {et~al.}(2003)\citenamefont
  {Manoharan}, \citenamefont {Elsesser},\ and\ \citenamefont
  {Pine}}]{manoharan2003dense}%
  \BibitemOpen
  \bibfield  {author} {\bibinfo {author} {\bibfnamefont {V.~N.}\ \bibnamefont
  {Manoharan}}, \bibinfo {author} {\bibfnamefont {M.~T.}\ \bibnamefont
  {Elsesser}},\ and\ \bibinfo {author} {\bibfnamefont {D.~J.}\ \bibnamefont
  {Pine}},\ }\bibfield  {title} {\bibinfo {title} {Dense packing and symmetry
  in small clusters of microspheres},\ }\href@noop {} {\bibfield  {journal}
  {\bibinfo  {journal} {Science}\ }\textbf {\bibinfo {volume} {301}},\ \bibinfo
  {pages} {483} (\bibinfo {year} {2003})}\BibitemShut {NoStop}%
\bibitem [{\citenamefont {Zhang}\ \emph {et~al.}(2005)\citenamefont {Zhang},
  \citenamefont {Keys}, \citenamefont {Chen},\ and\ \citenamefont
  {Glotzer}}]{zhang2005self}%
  \BibitemOpen
  \bibfield  {author} {\bibinfo {author} {\bibfnamefont {Z.}~\bibnamefont
  {Zhang}}, \bibinfo {author} {\bibfnamefont {A.~S.}\ \bibnamefont {Keys}},
  \bibinfo {author} {\bibfnamefont {T.}~\bibnamefont {Chen}},\ and\ \bibinfo
  {author} {\bibfnamefont {S.~C.}\ \bibnamefont {Glotzer}},\ }\bibfield
  {title} {\bibinfo {title} {Self-assembly of patchy particles into diamond
  structures through molecular mimicry},\ }\href@noop {} {\bibfield  {journal}
  {\bibinfo  {journal} {Langmuir}\ }\textbf {\bibinfo {volume} {21}},\ \bibinfo
  {pages} {11547} (\bibinfo {year} {2005})}\BibitemShut {NoStop}%
\bibitem [{\citenamefont {Romano}\ \emph {et~al.}(2014)\citenamefont {Romano},
  \citenamefont {Russo},\ and\ \citenamefont {Tanaka}}]{romano2014influence}%
  \BibitemOpen
  \bibfield  {author} {\bibinfo {author} {\bibfnamefont {F.}~\bibnamefont
  {Romano}}, \bibinfo {author} {\bibfnamefont {J.}~\bibnamefont {Russo}},\ and\
  \bibinfo {author} {\bibfnamefont {H.}~\bibnamefont {Tanaka}},\ }\bibfield
  {title} {\bibinfo {title} {Influence of patch-size variability on the
  crystallization of tetrahedral patchy particles},\ }\href@noop {} {\bibfield
  {journal} {\bibinfo  {journal} {Physical review letters}\ }\textbf {\bibinfo
  {volume} {113}},\ \bibinfo {pages} {138303} (\bibinfo {year}
  {2014})}\BibitemShut {NoStop}%
\bibitem [{\citenamefont {Halverson}\ and\ \citenamefont
  {Tkachenko}(2013)}]{halverson2013dna}%
  \BibitemOpen
  \bibfield  {author} {\bibinfo {author} {\bibfnamefont {J.~D.}\ \bibnamefont
  {Halverson}}\ and\ \bibinfo {author} {\bibfnamefont {A.~V.}\ \bibnamefont
  {Tkachenko}},\ }\bibfield  {title} {\bibinfo {title} {Dna-programmed
  mesoscopic architecture},\ }\href@noop {} {\bibfield  {journal} {\bibinfo
  {journal} {Physical Review E}\ }\textbf {\bibinfo {volume} {87}},\ \bibinfo
  {pages} {062310} (\bibinfo {year} {2013})}\BibitemShut {NoStop}%
\bibitem [{\citenamefont {Romano}\ and\ \citenamefont
  {Sciortino}(2012)}]{romano2012patterning}%
  \BibitemOpen
  \bibfield  {author} {\bibinfo {author} {\bibfnamefont {F.}~\bibnamefont
  {Romano}}\ and\ \bibinfo {author} {\bibfnamefont {F.}~\bibnamefont
  {Sciortino}},\ }\bibfield  {title} {\bibinfo {title} {Patterning symmetry in
  the rational design of colloidal crystals},\ }\href@noop {} {\bibfield
  {journal} {\bibinfo  {journal} {Nature communications}\ }\textbf {\bibinfo
  {volume} {3}},\ \bibinfo {pages} {975} (\bibinfo {year} {2012})}\BibitemShut
  {NoStop}%
\bibitem [{\citenamefont {Tracey}\ \emph {et~al.}(2019)\citenamefont {Tracey},
  \citenamefont {Noya},\ and\ \citenamefont {Doye}}]{tracey2019programming}%
  \BibitemOpen
  \bibfield  {author} {\bibinfo {author} {\bibfnamefont {D.~F.}\ \bibnamefont
  {Tracey}}, \bibinfo {author} {\bibfnamefont {E.~G.}\ \bibnamefont {Noya}},\
  and\ \bibinfo {author} {\bibfnamefont {J.~P.~K.}\ \bibnamefont {Doye}},\
  }\bibfield  {title} {\bibinfo {title} {Programming patchy particles to form
  complex periodic structures},\ }\href@noop {} {\bibfield  {journal} {\bibinfo
   {journal} {The Journal of Chemical Physics}\ }\textbf {\bibinfo {volume}
  {151}},\ \bibinfo {pages} {224506} (\bibinfo {year} {2019})}\BibitemShut
  {NoStop}%
\bibitem [{\citenamefont {Romano}\ \emph {et~al.}(2020)\citenamefont {Romano},
  \citenamefont {Russo}, \citenamefont {Kroc},\ and\ \citenamefont
  {{\v{S}}ulc}}]{romano2020designing}%
  \BibitemOpen
  \bibfield  {author} {\bibinfo {author} {\bibfnamefont {F.}~\bibnamefont
  {Romano}}, \bibinfo {author} {\bibfnamefont {J.}~\bibnamefont {Russo}},
  \bibinfo {author} {\bibfnamefont {L.}~\bibnamefont {Kroc}},\ and\ \bibinfo
  {author} {\bibfnamefont {P.}~\bibnamefont {{\v{S}}ulc}},\ }\bibfield  {title}
  {\bibinfo {title} {Designing patchy interactions to self-assemble arbitrary
  structures},\ }\href@noop {} {\bibfield  {journal} {\bibinfo  {journal}
  {Physical Review Letters}\ }\textbf {\bibinfo {volume} {125}},\ \bibinfo
  {pages} {118003} (\bibinfo {year} {2020})}\BibitemShut {NoStop}%
\bibitem [{\citenamefont {Russo}\ \emph {et~al.}(2022)\citenamefont {Russo},
  \citenamefont {Romano}, \citenamefont {Kroc}, \citenamefont {Sciortino},
  \citenamefont {Rovigatti},\ and\ \citenamefont {{\v{S}}ulc}}]{russo2022sat}%
  \BibitemOpen
  \bibfield  {author} {\bibinfo {author} {\bibfnamefont {J.}~\bibnamefont
  {Russo}}, \bibinfo {author} {\bibfnamefont {F.}~\bibnamefont {Romano}},
  \bibinfo {author} {\bibfnamefont {L.}~\bibnamefont {Kroc}}, \bibinfo {author}
  {\bibfnamefont {F.}~\bibnamefont {Sciortino}}, \bibinfo {author}
  {\bibfnamefont {L.}~\bibnamefont {Rovigatti}},\ and\ \bibinfo {author}
  {\bibfnamefont {P.}~\bibnamefont {{\v{S}}ulc}},\ }\bibfield  {title}
  {\bibinfo {title} {Sat-assembly: A new approach for designing self-assembling
  systems},\ }\href@noop {} {\bibfield  {journal} {\bibinfo  {journal} {Journal
  of Physics: Condensed Matter}\ } (\bibinfo {year} {2022})}\BibitemShut
  {NoStop}%
\bibitem [{\citenamefont {Rovigatti}\ \emph {et~al.}(2022)\citenamefont
  {Rovigatti}, \citenamefont {Russo}, \citenamefont {Romano}, \citenamefont
  {Matthies}, \citenamefont {Kroc},\ and\ \citenamefont
  {{\v{S}}ulc}}]{rovigatti2022simple}%
  \BibitemOpen
  \bibfield  {author} {\bibinfo {author} {\bibfnamefont {L.}~\bibnamefont
  {Rovigatti}}, \bibinfo {author} {\bibfnamefont {J.}~\bibnamefont {Russo}},
  \bibinfo {author} {\bibfnamefont {F.}~\bibnamefont {Romano}}, \bibinfo
  {author} {\bibfnamefont {M.}~\bibnamefont {Matthies}}, \bibinfo {author}
  {\bibfnamefont {L.}~\bibnamefont {Kroc}},\ and\ \bibinfo {author}
  {\bibfnamefont {P.}~\bibnamefont {{\v{S}}ulc}},\ }\bibfield  {title}
  {\bibinfo {title} {A simple solution to the problem of self-assembling cubic
  diamond crystals},\ }\href@noop {} {\bibfield  {journal} {\bibinfo  {journal}
  {Nanoscale}\ }\textbf {\bibinfo {volume} {14}},\ \bibinfo {pages} {14268}
  (\bibinfo {year} {2022})}\BibitemShut {NoStop}%
\bibitem [{\citenamefont {Desgranges}\ and\ \citenamefont
  {Delhommelle}(2014)}]{desgranges2014unraveling}%
  \BibitemOpen
  \bibfield  {author} {\bibinfo {author} {\bibfnamefont {C.}~\bibnamefont
  {Desgranges}}\ and\ \bibinfo {author} {\bibfnamefont {J.}~\bibnamefont
  {Delhommelle}},\ }\bibfield  {title} {\bibinfo {title} {Unraveling the
  coupling between demixing and crystallization in mixtures},\ }\href@noop {}
  {\bibfield  {journal} {\bibinfo  {journal} {Journal of the American Chemical
  Society}\ }\textbf {\bibinfo {volume} {136}},\ \bibinfo {pages} {8145}
  (\bibinfo {year} {2014})}\BibitemShut {NoStop}%
\bibitem [{\citenamefont {Wang}\ \emph {et~al.}(2010)\citenamefont {Wang},
  \citenamefont {Lomakin}, \citenamefont {McManus}, \citenamefont {Ogun},\ and\
  \citenamefont {Benedek}}]{wang2010phase}%
  \BibitemOpen
  \bibfield  {author} {\bibinfo {author} {\bibfnamefont {Y.}~\bibnamefont
  {Wang}}, \bibinfo {author} {\bibfnamefont {A.}~\bibnamefont {Lomakin}},
  \bibinfo {author} {\bibfnamefont {J.~J.}\ \bibnamefont {McManus}}, \bibinfo
  {author} {\bibfnamefont {O.}~\bibnamefont {Ogun}},\ and\ \bibinfo {author}
  {\bibfnamefont {G.~B.}\ \bibnamefont {Benedek}},\ }\bibfield  {title}
  {\bibinfo {title} {Phase behavior of mixtures of human lens proteins gamma d
  and beta b1},\ }\href@noop {} {\bibfield  {journal} {\bibinfo  {journal}
  {Proceedings of the National Academy of Sciences}\ }\textbf {\bibinfo
  {volume} {107}},\ \bibinfo {pages} {13282} (\bibinfo {year}
  {2010})}\BibitemShut {NoStop}%
\bibitem [{\citenamefont {Wang}\ \emph {et~al.}(2011)\citenamefont {Wang},
  \citenamefont {Lomakin}, \citenamefont {Latypov},\ and\ \citenamefont
  {Benedek}}]{wang2011phase}%
  \BibitemOpen
  \bibfield  {author} {\bibinfo {author} {\bibfnamefont {Y.}~\bibnamefont
  {Wang}}, \bibinfo {author} {\bibfnamefont {A.}~\bibnamefont {Lomakin}},
  \bibinfo {author} {\bibfnamefont {R.~F.}\ \bibnamefont {Latypov}},\ and\
  \bibinfo {author} {\bibfnamefont {G.~B.}\ \bibnamefont {Benedek}},\
  }\bibfield  {title} {\bibinfo {title} {Phase separation in solutions of
  monoclonal antibodies and the effect of human serum albumin},\ }\href@noop {}
  {\bibfield  {journal} {\bibinfo  {journal} {Proceedings of the National
  Academy of Sciences}\ }\textbf {\bibinfo {volume} {108}},\ \bibinfo {pages}
  {16606} (\bibinfo {year} {2011})}\BibitemShut {NoStop}%
\bibitem [{\citenamefont {Heidenreich}\ \emph {et~al.}(2020)\citenamefont
  {Heidenreich}, \citenamefont {Georgeson}, \citenamefont {Locatelli},
  \citenamefont {Rovigatti}, \citenamefont {Nandi}, \citenamefont {Steinberg},
  \citenamefont {Nadav}, \citenamefont {Shimoni}, \citenamefont {Safran},
  \citenamefont {Doye},\ and\ \citenamefont {Levy}}]{heidenreich2020designer}%
  \BibitemOpen
  \bibfield  {author} {\bibinfo {author} {\bibfnamefont {M.}~\bibnamefont
  {Heidenreich}}, \bibinfo {author} {\bibfnamefont {J.~M.}\ \bibnamefont
  {Georgeson}}, \bibinfo {author} {\bibfnamefont {E.}~\bibnamefont
  {Locatelli}}, \bibinfo {author} {\bibfnamefont {L.}~\bibnamefont
  {Rovigatti}}, \bibinfo {author} {\bibfnamefont {S.~K.}\ \bibnamefont
  {Nandi}}, \bibinfo {author} {\bibfnamefont {A.}~\bibnamefont {Steinberg}},
  \bibinfo {author} {\bibfnamefont {Y.}~\bibnamefont {Nadav}}, \bibinfo
  {author} {\bibfnamefont {E.}~\bibnamefont {Shimoni}}, \bibinfo {author}
  {\bibfnamefont {S.~A.}\ \bibnamefont {Safran}}, \bibinfo {author}
  {\bibfnamefont {J.~P.}\ \bibnamefont {Doye}},\ and\ \bibinfo {author}
  {\bibfnamefont {E.~D.}\ \bibnamefont {Levy}},\ }\bibfield  {title} {\bibinfo
  {title} {Designer protein assemblies with tunable phase diagrams in living
  cells},\ }\href@noop {} {\bibfield  {journal} {\bibinfo  {journal} {Nature
  Chemical Biology}\ }\textbf {\bibinfo {volume} {16}},\ \bibinfo {pages} {939}
  (\bibinfo {year} {2020})}\BibitemShut {NoStop}%
\bibitem [{\citenamefont {van Anders}\ \emph {et~al.}(2013)\citenamefont {van
  Anders}, \citenamefont {Ahmed}, \citenamefont {Smith}, \citenamefont
  {Engel},\ and\ \citenamefont {Glotzer}}]{van2013entropically}%
  \BibitemOpen
  \bibfield  {author} {\bibinfo {author} {\bibfnamefont {G.}~\bibnamefont {van
  Anders}}, \bibinfo {author} {\bibfnamefont {N.~K.}\ \bibnamefont {Ahmed}},
  \bibinfo {author} {\bibfnamefont {R.}~\bibnamefont {Smith}}, \bibinfo
  {author} {\bibfnamefont {M.}~\bibnamefont {Engel}},\ and\ \bibinfo {author}
  {\bibfnamefont {S.~C.}\ \bibnamefont {Glotzer}},\ }\bibfield  {title}
  {\bibinfo {title} {Entropically patchy particles: engineering valence through
  shape entropy},\ }\href@noop {} {\bibfield  {journal} {\bibinfo  {journal}
  {Acs Nano}\ }\textbf {\bibinfo {volume} {8}},\ \bibinfo {pages} {931}
  (\bibinfo {year} {2013})}\BibitemShut {NoStop}%
\bibitem [{\citenamefont {Zhang}\ and\ \citenamefont
  {Glotzer}(2004)}]{zhang2004self}%
  \BibitemOpen
  \bibfield  {author} {\bibinfo {author} {\bibfnamefont {Z.}~\bibnamefont
  {Zhang}}\ and\ \bibinfo {author} {\bibfnamefont {S.~C.}\ \bibnamefont
  {Glotzer}},\ }\bibfield  {title} {\bibinfo {title} {Self-assembly of patchy
  particles},\ }\href@noop {} {\bibfield  {journal} {\bibinfo  {journal} {Nano
  Letters}\ }\textbf {\bibinfo {volume} {4}},\ \bibinfo {pages} {1407}
  (\bibinfo {year} {2004})}\BibitemShut {NoStop}%
\bibitem [{\citenamefont {Pawar}\ and\ \citenamefont
  {Kretzschmar}(2010)}]{pawar2010fabrication}%
  \BibitemOpen
  \bibfield  {author} {\bibinfo {author} {\bibfnamefont {A.~B.}\ \bibnamefont
  {Pawar}}\ and\ \bibinfo {author} {\bibfnamefont {I.}~\bibnamefont
  {Kretzschmar}},\ }\bibfield  {title} {\bibinfo {title} {Fabrication,
  assembly, and application of patchy particles},\ }\href@noop {} {\bibfield
  {journal} {\bibinfo  {journal} {Macromolecular rapid communications}\
  }\textbf {\bibinfo {volume} {31}},\ \bibinfo {pages} {150} (\bibinfo {year}
  {2010})}\BibitemShut {NoStop}%
\bibitem [{\citenamefont {Bianchi}\ \emph {et~al.}(2011)\citenamefont
  {Bianchi}, \citenamefont {Blaak},\ and\ \citenamefont
  {Likos}}]{bianchi2011patchy}%
  \BibitemOpen
  \bibfield  {author} {\bibinfo {author} {\bibfnamefont {E.}~\bibnamefont
  {Bianchi}}, \bibinfo {author} {\bibfnamefont {R.}~\bibnamefont {Blaak}},\
  and\ \bibinfo {author} {\bibfnamefont {C.~N.}\ \bibnamefont {Likos}},\
  }\bibfield  {title} {\bibinfo {title} {Patchy colloids: state of the art and
  perspectives},\ }\href@noop {} {\bibfield  {journal} {\bibinfo  {journal}
  {Physical Chemistry Chemical Physics}\ }\textbf {\bibinfo {volume} {13}},\
  \bibinfo {pages} {6397} (\bibinfo {year} {2011})}\BibitemShut {NoStop}%
\bibitem [{\citenamefont {Romano}\ and\ \citenamefont
  {Sciortino}(2011)}]{romano2011colloidal}%
  \BibitemOpen
  \bibfield  {author} {\bibinfo {author} {\bibfnamefont {F.}~\bibnamefont
  {Romano}}\ and\ \bibinfo {author} {\bibfnamefont {F.}~\bibnamefont
  {Sciortino}},\ }\bibfield  {title} {\bibinfo {title} {Colloidal
  self-assembly: patchy from the bottom up},\ }\href@noop {} {\bibfield
  {journal} {\bibinfo  {journal} {Nature materials}\ }\textbf {\bibinfo
  {volume} {10}},\ \bibinfo {pages} {171} (\bibinfo {year} {2011})}\BibitemShut
  {NoStop}%
\bibitem [{\citenamefont {Suzuki}\ \emph {et~al.}(2009)\citenamefont {Suzuki},
  \citenamefont {Hosokawa},\ and\ \citenamefont
  {Maeda}}]{suzuki2009controlling}%
  \BibitemOpen
  \bibfield  {author} {\bibinfo {author} {\bibfnamefont {K.}~\bibnamefont
  {Suzuki}}, \bibinfo {author} {\bibfnamefont {K.}~\bibnamefont {Hosokawa}},\
  and\ \bibinfo {author} {\bibfnamefont {M.}~\bibnamefont {Maeda}},\ }\bibfield
   {title} {\bibinfo {title} {Controlling the number and positions of
  oligonucleotides on gold nanoparticle surfaces},\ }\href@noop {} {\bibfield
  {journal} {\bibinfo  {journal} {Journal of the American Chemical Society}\
  }\textbf {\bibinfo {volume} {131}},\ \bibinfo {pages} {7518} (\bibinfo {year}
  {2009})}\BibitemShut {NoStop}%
\bibitem [{\citenamefont {Kim}\ \emph {et~al.}(2011)\citenamefont {Kim},
  \citenamefont {Kim},\ and\ \citenamefont {Deaton}}]{kim2011dna}%
  \BibitemOpen
  \bibfield  {author} {\bibinfo {author} {\bibfnamefont {J.-W.}\ \bibnamefont
  {Kim}}, \bibinfo {author} {\bibfnamefont {J.-H.}\ \bibnamefont {Kim}},\ and\
  \bibinfo {author} {\bibfnamefont {R.}~\bibnamefont {Deaton}},\ }\bibfield
  {title} {\bibinfo {title} {Dna-linked nanoparticle building blocks for
  programmable matter},\ }\href@noop {} {\bibfield  {journal} {\bibinfo
  {journal} {Angewandte Chemie International Edition}\ }\textbf {\bibinfo
  {volume} {50}},\ \bibinfo {pages} {9185} (\bibinfo {year}
  {2011})}\BibitemShut {NoStop}%
\bibitem [{\citenamefont {Wang}\ \emph {et~al.}(2012)\citenamefont {Wang},
  \citenamefont {Wang}, \citenamefont {Breed}, \citenamefont {Manoharan},
  \citenamefont {Feng}, \citenamefont {Hollingsworth}, \citenamefont {Weck},\
  and\ \citenamefont {Pine}}]{wang2012colloids}%
  \BibitemOpen
  \bibfield  {author} {\bibinfo {author} {\bibfnamefont {Y.}~\bibnamefont
  {Wang}}, \bibinfo {author} {\bibfnamefont {Y.}~\bibnamefont {Wang}}, \bibinfo
  {author} {\bibfnamefont {D.~R.}\ \bibnamefont {Breed}}, \bibinfo {author}
  {\bibfnamefont {V.~N.}\ \bibnamefont {Manoharan}}, \bibinfo {author}
  {\bibfnamefont {L.}~\bibnamefont {Feng}}, \bibinfo {author} {\bibfnamefont
  {A.~D.}\ \bibnamefont {Hollingsworth}}, \bibinfo {author} {\bibfnamefont
  {M.}~\bibnamefont {Weck}},\ and\ \bibinfo {author} {\bibfnamefont {D.~J.}\
  \bibnamefont {Pine}},\ }\bibfield  {title} {\bibinfo {title} {Colloids with
  valence and specific directional bonding},\ }\href@noop {} {\bibfield
  {journal} {\bibinfo  {journal} {Nature}\ }\textbf {\bibinfo {volume} {491}},\
  \bibinfo {pages} {51} (\bibinfo {year} {2012})}\BibitemShut {NoStop}%
\bibitem [{\citenamefont {Feng}\ \emph {et~al.}(2013)\citenamefont {Feng},
  \citenamefont {Dreyfus}, \citenamefont {Sha}, \citenamefont {Seeman},\ and\
  \citenamefont {Chaikin}}]{feng2013dna}%
  \BibitemOpen
  \bibfield  {author} {\bibinfo {author} {\bibfnamefont {L.}~\bibnamefont
  {Feng}}, \bibinfo {author} {\bibfnamefont {R.}~\bibnamefont {Dreyfus}},
  \bibinfo {author} {\bibfnamefont {R.}~\bibnamefont {Sha}}, \bibinfo {author}
  {\bibfnamefont {N.~C.}\ \bibnamefont {Seeman}},\ and\ \bibinfo {author}
  {\bibfnamefont {P.~M.}\ \bibnamefont {Chaikin}},\ }\bibfield  {title}
  {\bibinfo {title} {Dna patchy particles},\ }\href@noop {} {\bibfield
  {journal} {\bibinfo  {journal} {Advanced Materials}\ }\textbf {\bibinfo
  {volume} {25}},\ \bibinfo {pages} {2779} (\bibinfo {year}
  {2013})}\BibitemShut {NoStop}%
\bibitem [{\citenamefont {Rothemund}(2006)}]{Rothemund2006}%
  \BibitemOpen
  \bibfield  {author} {\bibinfo {author} {\bibfnamefont {P.~W.~K.}\
  \bibnamefont {Rothemund}},\ }\bibfield  {title} {\bibinfo {title} {{Folding
  DNA to create nanoscale shapes and patterns}},\ }\href
  {https://doi.org/10.1038/nature04586} {\bibfield  {journal} {\bibinfo
  {journal} {Nature}\ }\textbf {\bibinfo {volume} {440}},\ \bibinfo {pages}
  {297} (\bibinfo {year} {2006})}\BibitemShut {NoStop}%
\bibitem [{\citenamefont {Tian}\ \emph {et~al.}(2020)\citenamefont {Tian},
  \citenamefont {Lhermitte}, \citenamefont {Bai}, \citenamefont {Vo},
  \citenamefont {Xin}, \citenamefont {Li}, \citenamefont {Li}, \citenamefont
  {Fukuto}, \citenamefont {Yager}, \citenamefont {Kahn} \emph
  {et~al.}}]{tian2020ordered}%
  \BibitemOpen
  \bibfield  {author} {\bibinfo {author} {\bibfnamefont {Y.}~\bibnamefont
  {Tian}}, \bibinfo {author} {\bibfnamefont {J.~R.}\ \bibnamefont {Lhermitte}},
  \bibinfo {author} {\bibfnamefont {L.}~\bibnamefont {Bai}}, \bibinfo {author}
  {\bibfnamefont {T.}~\bibnamefont {Vo}}, \bibinfo {author} {\bibfnamefont
  {H.~L.}\ \bibnamefont {Xin}}, \bibinfo {author} {\bibfnamefont
  {H.}~\bibnamefont {Li}}, \bibinfo {author} {\bibfnamefont {R.}~\bibnamefont
  {Li}}, \bibinfo {author} {\bibfnamefont {M.}~\bibnamefont {Fukuto}}, \bibinfo
  {author} {\bibfnamefont {K.~G.}\ \bibnamefont {Yager}}, \bibinfo {author}
  {\bibfnamefont {J.~S.}\ \bibnamefont {Kahn}}, \emph {et~al.},\ }\bibfield
  {title} {\bibinfo {title} {Ordered three-dimensional nanomaterials using
  dna-prescribed and valence-controlled material voxels},\ }\href@noop {}
  {\bibfield  {journal} {\bibinfo  {journal} {Nature materials}\ }\textbf
  {\bibinfo {volume} {19}},\ \bibinfo {pages} {789} (\bibinfo {year}
  {2020})}\BibitemShut {NoStop}%
\bibitem [{\citenamefont {Bol}(1982)}]{bol1982monte}%
  \BibitemOpen
  \bibfield  {author} {\bibinfo {author} {\bibfnamefont {W.}~\bibnamefont
  {Bol}},\ }\bibfield  {title} {\bibinfo {title} {Monte {C}arlo simulations of
  fluid systems of waterlike molecules},\ }\href@noop {} {\bibfield  {journal}
  {\bibinfo  {journal} {Molecular Physics}\ }\textbf {\bibinfo {volume} {45}},\
  \bibinfo {pages} {605} (\bibinfo {year} {1982})}\BibitemShut {NoStop}%
\bibitem [{\citenamefont {Kern}\ and\ \citenamefont
  {Frenkel}(2003)}]{kern2003fluid}%
  \BibitemOpen
  \bibfield  {author} {\bibinfo {author} {\bibfnamefont {N.}~\bibnamefont
  {Kern}}\ and\ \bibinfo {author} {\bibfnamefont {D.}~\bibnamefont {Frenkel}},\
  }\bibfield  {title} {\bibinfo {title} {Fluid--fluid coexistence in colloidal
  systems with short-ranged strongly directional attraction},\ }\href@noop {}
  {\bibfield  {journal} {\bibinfo  {journal} {The Journal of chemical physics}\
  }\textbf {\bibinfo {volume} {118}},\ \bibinfo {pages} {9882} (\bibinfo {year}
  {2003})}\BibitemShut {NoStop}%
\bibitem [{\citenamefont {Rovigatti}\ \emph {et~al.}(2018)\citenamefont
  {Rovigatti}, \citenamefont {Russo},\ and\ \citenamefont
  {Romano}}]{rovigatti2018simulate}%
  \BibitemOpen
  \bibfield  {author} {\bibinfo {author} {\bibfnamefont {L.}~\bibnamefont
  {Rovigatti}}, \bibinfo {author} {\bibfnamefont {J.}~\bibnamefont {Russo}},\
  and\ \bibinfo {author} {\bibfnamefont {F.}~\bibnamefont {Romano}},\
  }\bibfield  {title} {\bibinfo {title} {How to simulate patchy particles},\
  }\href@noop {} {\bibfield  {journal} {\bibinfo  {journal} {The European
  Physical Journal E}\ }\textbf {\bibinfo {volume} {41}},\ \bibinfo {pages}
  {59} (\bibinfo {year} {2018})}\BibitemShut {NoStop}%
\bibitem [{\citenamefont {Beneduce}\ \emph {et~al.}(2022)\citenamefont
  {Beneduce}, \citenamefont {Sciortino}, \citenamefont {Sulc},\ and\
  \citenamefont {Russo}}]{beneduce2022include}%
  \BibitemOpen
  \bibfield  {author} {\bibinfo {author} {\bibfnamefont {C.}~\bibnamefont
  {Beneduce}}, \bibinfo {author} {\bibfnamefont {F.}~\bibnamefont {Sciortino}},
  \bibinfo {author} {\bibfnamefont {P.}~\bibnamefont {Sulc}},\ and\ \bibinfo
  {author} {\bibfnamefont {J.}~\bibnamefont {Russo}},\ }\bibfield  {title}
  {\bibinfo {title} {How to include azeotropy in the design of self-assembling
  patchy particles systems},\ }\href@noop {} {\bibfield  {journal} {\bibinfo
  {journal} {arXiv preprint arXiv:2208.09856}\ } (\bibinfo {year}
  {2022})}\BibitemShut {NoStop}%
\bibitem [{\citenamefont {Rubinstein}\ \emph {et~al.}(2003)\citenamefont
  {Rubinstein}, \citenamefont {Colby} \emph {et~al.}}]{rubinstein2003polymer}%
  \BibitemOpen
  \bibfield  {author} {\bibinfo {author} {\bibfnamefont {M.}~\bibnamefont
  {Rubinstein}}, \bibinfo {author} {\bibfnamefont {R.~H.}\ \bibnamefont
  {Colby}}, \emph {et~al.},\ }\href@noop {} {\emph {\bibinfo {title} {Polymer
  physics}}},\ Vol.~\bibinfo {volume} {23}\ (\bibinfo  {publisher} {Oxford
  university press New York},\ \bibinfo {year} {2003})\BibitemShut {NoStop}%
\bibitem [{\citenamefont {Tanaka}\ \emph {et~al.}(2019)\citenamefont {Tanaka},
  \citenamefont {Tong}, \citenamefont {Shi},\ and\ \citenamefont
  {Russo}}]{tanaka2019revealing}%
  \BibitemOpen
  \bibfield  {author} {\bibinfo {author} {\bibfnamefont {H.}~\bibnamefont
  {Tanaka}}, \bibinfo {author} {\bibfnamefont {H.}~\bibnamefont {Tong}},
  \bibinfo {author} {\bibfnamefont {R.}~\bibnamefont {Shi}},\ and\ \bibinfo
  {author} {\bibfnamefont {J.}~\bibnamefont {Russo}},\ }\bibfield  {title}
  {\bibinfo {title} {Revealing key structural features hidden in liquids and
  glasses},\ }\href@noop {} {\bibfield  {journal} {\bibinfo  {journal} {Nature
  Reviews Physics}\ }\textbf {\bibinfo {volume} {1}},\ \bibinfo {pages} {333}
  (\bibinfo {year} {2019})}\BibitemShut {NoStop}%
\bibitem [{\citenamefont {Lechner}\ and\ \citenamefont
  {Dellago}(2008)}]{lechner2008accurate}%
  \BibitemOpen
  \bibfield  {author} {\bibinfo {author} {\bibfnamefont {W.}~\bibnamefont
  {Lechner}}\ and\ \bibinfo {author} {\bibfnamefont {C.}~\bibnamefont
  {Dellago}},\ }\bibfield  {title} {\bibinfo {title} {Accurate determination of
  crystal structures based on averaged local bond order parameters},\
  }\href@noop {} {\bibfield  {journal} {\bibinfo  {journal} {The Journal of
  chemical physics}\ }\textbf {\bibinfo {volume} {129}},\ \bibinfo {pages}
  {114707} (\bibinfo {year} {2008})}\BibitemShut {NoStop}%
\bibitem [{\citenamefont {Frenkel}\ and\ \citenamefont
  {Smit}(2001)}]{frenkel2001understanding}%
  \BibitemOpen
  \bibfield  {author} {\bibinfo {author} {\bibfnamefont {D.}~\bibnamefont
  {Frenkel}}\ and\ \bibinfo {author} {\bibfnamefont {B.}~\bibnamefont {Smit}},\
  }\href@noop {} {\emph {\bibinfo {title} {Understanding molecular simulation:
  from algorithms to applications}}},\ Vol.~\bibinfo {volume} {1}\ (\bibinfo
  {publisher} {Elsevier},\ \bibinfo {year} {2001})\BibitemShut {NoStop}%
\bibitem [{\citenamefont {Xu}\ \emph {et~al.}(2012)\citenamefont {Xu},
  \citenamefont {Buldyrev}, \citenamefont {Stanley},\ and\ \citenamefont
  {Franzese}}]{xu2012homogeneous}%
  \BibitemOpen
  \bibfield  {author} {\bibinfo {author} {\bibfnamefont {L.}~\bibnamefont
  {Xu}}, \bibinfo {author} {\bibfnamefont {S.~V.}\ \bibnamefont {Buldyrev}},
  \bibinfo {author} {\bibfnamefont {H.~E.}\ \bibnamefont {Stanley}},\ and\
  \bibinfo {author} {\bibfnamefont {G.}~\bibnamefont {Franzese}},\ }\bibfield
  {title} {\bibinfo {title} {Homogeneous crystal nucleation near a metastable
  fluid-fluid phase transition},\ }\href@noop {} {\bibfield  {journal}
  {\bibinfo  {journal} {Physical Review Letters}\ }\textbf {\bibinfo {volume}
  {109}},\ \bibinfo {pages} {095702} (\bibinfo {year} {2012})}\BibitemShut
  {NoStop}%
\bibitem [{\citenamefont {Stukowski}(2010)}]{ovito}%
  \BibitemOpen
  \bibfield  {author} {\bibinfo {author} {\bibfnamefont {A.}~\bibnamefont
  {Stukowski}},\ }\bibfield  {title} {\bibinfo {title} {{Visualization and
  analysis of atomistic simulation data with OVITO-the Open Visualization
  Tool}},\ }\bibfield  {journal} {\bibinfo  {journal} {{MODELLING AND
  SIMULATION IN MATERIALS SCIENCE AND ENGINEERING}}\ }\textbf {\bibinfo
  {volume} {{18}}},\ \href {https://doi.org/{10.1088/0965-0393/18/1/015012}}
  {{10.1088/0965-0393/18/1/015012}} (\bibinfo {year} {{2010}})\BibitemShut
  {NoStop}%
\bibitem [{\citenamefont {Chakraborty}\ \emph {et~al.}(2022)\citenamefont
  {Chakraborty}, \citenamefont {Pearce}, \citenamefont {Verweij}, \citenamefont
  {Matysik}, \citenamefont {Giomi},\ and\ \citenamefont
  {Kraft}}]{chakraborty2022self}%
  \BibitemOpen
  \bibfield  {author} {\bibinfo {author} {\bibfnamefont {I.}~\bibnamefont
  {Chakraborty}}, \bibinfo {author} {\bibfnamefont {D.~J.}\ \bibnamefont
  {Pearce}}, \bibinfo {author} {\bibfnamefont {R.~W.}\ \bibnamefont {Verweij}},
  \bibinfo {author} {\bibfnamefont {S.~C.}\ \bibnamefont {Matysik}}, \bibinfo
  {author} {\bibfnamefont {L.}~\bibnamefont {Giomi}},\ and\ \bibinfo {author}
  {\bibfnamefont {D.~J.}\ \bibnamefont {Kraft}},\ }\bibfield  {title} {\bibinfo
  {title} {Self-assembly dynamics of reconfigurable colloidal molecules},\
  }\href@noop {} {\bibfield  {journal} {\bibinfo  {journal} {ACS nano}\
  }\textbf {\bibinfo {volume} {16}},\ \bibinfo {pages} {2471} (\bibinfo {year}
  {2022})}\BibitemShut {NoStop}%
\bibitem [{\citenamefont {Xiong}\ \emph {et~al.}(2020)\citenamefont {Xiong},
  \citenamefont {Yang}, \citenamefont {Tian}, \citenamefont {Michelson},
  \citenamefont {Xiang}, \citenamefont {Xin},\ and\ \citenamefont
  {Gang}}]{xiong2020three}%
  \BibitemOpen
  \bibfield  {author} {\bibinfo {author} {\bibfnamefont {Y.}~\bibnamefont
  {Xiong}}, \bibinfo {author} {\bibfnamefont {S.}~\bibnamefont {Yang}},
  \bibinfo {author} {\bibfnamefont {Y.}~\bibnamefont {Tian}}, \bibinfo {author}
  {\bibfnamefont {A.}~\bibnamefont {Michelson}}, \bibinfo {author}
  {\bibfnamefont {S.}~\bibnamefont {Xiang}}, \bibinfo {author} {\bibfnamefont
  {H.}~\bibnamefont {Xin}},\ and\ \bibinfo {author} {\bibfnamefont
  {O.}~\bibnamefont {Gang}},\ }\bibfield  {title} {\bibinfo {title}
  {Three-dimensional patterning of nanoparticles by molecular stamping},\
  }\href@noop {} {\bibfield  {journal} {\bibinfo  {journal} {ACS nano}\
  }\textbf {\bibinfo {volume} {14}},\ \bibinfo {pages} {6823} (\bibinfo {year}
  {2020})}\BibitemShut {NoStop}%
\bibitem [{\citenamefont {Russo}\ \emph {et~al.}(2021)\citenamefont {Russo},
  \citenamefont {Leoni}, \citenamefont {Martelli},\ and\ \citenamefont
  {Sciortino}}]{russo2021physics}%
  \BibitemOpen
  \bibfield  {author} {\bibinfo {author} {\bibfnamefont {J.}~\bibnamefont
  {Russo}}, \bibinfo {author} {\bibfnamefont {F.}~\bibnamefont {Leoni}},
  \bibinfo {author} {\bibfnamefont {F.}~\bibnamefont {Martelli}},\ and\
  \bibinfo {author} {\bibfnamefont {F.}~\bibnamefont {Sciortino}},\ }\bibfield
  {title} {\bibinfo {title} {The physics of empty liquids: from patchy
  particles to water},\ }\href@noop {} {\bibfield  {journal} {\bibinfo
  {journal} {Reports on Progress in Physics}\ } (\bibinfo {year}
  {2021})}\BibitemShut {NoStop}%
\bibitem [{\citenamefont {Bianchi}\ \emph {et~al.}(2006)\citenamefont
  {Bianchi}, \citenamefont {Largo}, \citenamefont {Tartaglia}, \citenamefont
  {Zaccarelli},\ and\ \citenamefont {Sciortino}}]{bianchi2006empty}%
  \BibitemOpen
  \bibfield  {author} {\bibinfo {author} {\bibfnamefont {E.}~\bibnamefont
  {Bianchi}}, \bibinfo {author} {\bibfnamefont {J.}~\bibnamefont {Largo}},
  \bibinfo {author} {\bibfnamefont {P.}~\bibnamefont {Tartaglia}}, \bibinfo
  {author} {\bibfnamefont {E.}~\bibnamefont {Zaccarelli}},\ and\ \bibinfo
  {author} {\bibfnamefont {F.}~\bibnamefont {Sciortino}},\ }\bibfield  {title}
  {\bibinfo {title} {Phase diagram of patchy colloids: Towards empty liquids},\
  }\href@noop {} {\bibfield  {journal} {\bibinfo  {journal} {Physical Review
  Letters}\ }\textbf {\bibinfo {volume} {97}},\ \bibinfo {pages} {168301}
  (\bibinfo {year} {2006})}\BibitemShut {NoStop}%
\bibitem [{\citenamefont {Romano}\ \emph {et~al.}(2011)\citenamefont {Romano},
  \citenamefont {Sanz},\ and\ \citenamefont
  {Sciortino}}]{romano2011crystallization}%
  \BibitemOpen
  \bibfield  {author} {\bibinfo {author} {\bibfnamefont {F.}~\bibnamefont
  {Romano}}, \bibinfo {author} {\bibfnamefont {E.}~\bibnamefont {Sanz}},\ and\
  \bibinfo {author} {\bibfnamefont {F.}~\bibnamefont {Sciortino}},\ }\bibfield
  {title} {\bibinfo {title} {Crystallization of tetrahedral patchy particles in
  silico},\ }\href@noop {} {\bibfield  {journal} {\bibinfo  {journal} {The
  Journal of chemical physics}\ }\textbf {\bibinfo {volume} {134}},\ \bibinfo
  {pages} {174502} (\bibinfo {year} {2011})}\BibitemShut {NoStop}%
\bibitem [{\citenamefont {Altan}\ and\ \citenamefont
  {Charbonneau}(2019)}]{altan2019obtaining}%
  \BibitemOpen
  \bibfield  {author} {\bibinfo {author} {\bibfnamefont {I.}~\bibnamefont
  {Altan}}\ and\ \bibinfo {author} {\bibfnamefont {P.}~\bibnamefont
  {Charbonneau}},\ }\bibfield  {title} {\bibinfo {title} {Obtaining soft matter
  models of proteins and their phase behavior},\ }in\ \href@noop {} {\emph
  {\bibinfo {booktitle} {Protein Self-Assembly}}}\ (\bibinfo  {publisher}
  {Springer},\ \bibinfo {year} {2019})\ pp.\ \bibinfo {pages}
  {209--228}\BibitemShut {NoStop}%
\bibitem [{\citenamefont {Gnan}\ \emph {et~al.}(2019)\citenamefont {Gnan},
  \citenamefont {Sciortino},\ and\ \citenamefont
  {Zaccarelli}}]{gnan2019patchy}%
  \BibitemOpen
  \bibfield  {author} {\bibinfo {author} {\bibfnamefont {N.}~\bibnamefont
  {Gnan}}, \bibinfo {author} {\bibfnamefont {F.}~\bibnamefont {Sciortino}},\
  and\ \bibinfo {author} {\bibfnamefont {E.}~\bibnamefont {Zaccarelli}},\
  }\bibfield  {title} {\bibinfo {title} {Patchy particle models to understand
  protein phase behavior},\ }in\ \href@noop {} {\emph {\bibinfo {booktitle}
  {Protein Self-Assembly}}}\ (\bibinfo  {publisher} {Springer},\ \bibinfo
  {year} {2019})\ pp.\ \bibinfo {pages} {187--208}\BibitemShut {NoStop}%
\bibitem [{\citenamefont {McManus}(2019)}]{mcmanus2019protein}%
  \BibitemOpen
  \bibfield  {author} {\bibinfo {author} {\bibfnamefont {J.~J.}\ \bibnamefont
  {McManus}},\ }\href@noop {} {\emph {\bibinfo {title} {Protein
  Self-Assembly}}}\ (\bibinfo  {publisher} {Springer},\ \bibinfo {year}
  {2019})\BibitemShut {NoStop}%
\bibitem [{\citenamefont {Hvozd}\ \emph {et~al.}(2022)\citenamefont {Hvozd},
  \citenamefont {Kalyuzhnyi}, \citenamefont {Vlachy},\ and\ \citenamefont
  {Cummings}}]{hvozd2022empty}%
  \BibitemOpen
  \bibfield  {author} {\bibinfo {author} {\bibfnamefont {T.~V.}\ \bibnamefont
  {Hvozd}}, \bibinfo {author} {\bibfnamefont {Y.~V.}\ \bibnamefont
  {Kalyuzhnyi}}, \bibinfo {author} {\bibfnamefont {V.}~\bibnamefont {Vlachy}},\
  and\ \bibinfo {author} {\bibfnamefont {P.~T.}\ \bibnamefont {Cummings}},\
  }\bibfield  {title} {\bibinfo {title} {Empty liquid state and re-entrant
  phase behavior of the patchy colloids confined in porous media},\ }\href@noop
  {} {\bibfield  {journal} {\bibinfo  {journal} {The Journal of Chemical
  Physics}\ }\textbf {\bibinfo {volume} {156}},\ \bibinfo {pages} {161102}
  (\bibinfo {year} {2022})}\BibitemShut {NoStop}%
\end{thebibliography}
%

\end{document}